\documentclass[10pt,twocolumn,twoside]{IEEEtran}
\usepackage{amsmath}
\usepackage{mathtools, nccmath}
\usepackage{bbm}
\usepackage{amsthm}
\usepackage{amsfonts}
\usepackage{algorithm}
\usepackage{algorithmic}
\usepackage{verbatim}
\usepackage{url}
\def\BibTeX{{\rm B\kern-.05em{\sc i\kern-.025em b}\kern-.08em
    T\kern-.1667em\lower.7ex\hbox{E}\kern-.125emX}}

\newcommand{\vq}{\nu}

\newcommand{\xs}{\mathbf{q}}

\newcommand{\coeff}[1]{\frac{\delta}{#1 \alpha c_{(u,v)}}}
\newcommand\numberthis{\addtocounter{equation}{1}\tag{\theequation}}

\newcommand{\vlam}{\ensuremath{{\mbox{\boldmath{$\lambda$}}}} }
\usepackage[belowskip=-15pt,aboveskip=0pt]{caption}
\allowdisplaybreaks 
\usepackage{xcolor}
\usepackage{soul}
\newtheorem{definition}{Definition}[]
\newtheorem{theorem}{Theorem}[]
\newtheorem{lemma}{Lemma}[]
\newtheorem{proposition}{Proposition}[]

\newtheorem{assumption}{Assumption}[]

\ifCLASSINFOpdf
\else
\fi
\begin{document}
%
\title{Throughput Optimal Routing in Blockchain Based Payment Systems}

\author{\IEEEauthorblockN{Sushil Mahavir Varma, Siva Theja Maguluri  \\}
\IEEEauthorblockA{Industrial and Systems Engineering 
Georgia Institute of Technology
Atlanta, Georgia 30332 \\
Email: sushil@gatech.edu, siva.theja@gatech.edu}}


%


\maketitle

\begin{abstract}
 Cryptocurrency networks such as Bitcoin have emerged as a distributed alternative to traditional centralized financial transaction networks. However, there are major challenges in scaling up the throughput of such networks. Lightning network \cite{lightningnetwork} and Spider network \cite{spider} are alternates that build bidirectional payment channels on top of cryptocurrency networks using smart contracts, to enable fast transactions that bypass the Blockchain. 

In this paper, we study the problem of routing transactions in such a payment processing network. We first propose a Stochastic model to study such a system, as opposed to a fluid model that is studied in the literature. Each link in such a model is a two-sided queue, and unlike classical queues, such queues are not stable unless there is an external control. We propose a notion of stability for the payment processing network consisting of such two-sided queues using the notion of on-chain rebalancing. We then characterize the capacity region and propose a throughput optimal algorithm that stabilizes the system under any load within the capacity region. The stochastic model enables us to study closed loop policies, which typically have better queuing/delay performance than the open loop policies (or static split rules) studied in the literature. We investigate this through simulations.
\end{abstract}


%
\IEEEpeerreviewmaketitle

\section{Introduction}
Blockchain is a decentralized, distributed public ledger that is secured based on consensus and cryptographic hashing, and was first introduced in the seminal paper \cite{satoshibitcoin} to build the Bitcoin cryptocurrency Systems. Over the last decade, Blockchain has been used in a variety of applications, including to build other cryptocurrency networks \cite{wood2014ethereum}, Smart Contracts \cite{smartcontract,smartcontract2}, networks for financial services \cite{blockchainbanking}, supply chain management \cite{nash2016ibm} etc. In addition to the public blockchains such as Bitcoin and Ethereum, several companies are building private Blockchains \cite{marvin2017blockchain} and are also involved in Hybrid Blockchains \cite{hybridblockchain}. 

This paper studied payment transaction networks. In a centralized transaction network, a centralized body such as Paypal or Visa is responsible for facilitating all the transactions between any pair of sender and receiver. 
It is an efficient, fast and established payment system which achieves processing rate of thousand of transactions per second with only few seconds of delay. In such a network, each user of the system, has an account or some relation with the payment processor (possibly through another entity such as a bank). In contrast, in a decentralized transaction network, no such centralized entity is required. Credit networks such as \cite{ghosh2007mechanism} and Blockchain based public cryptocurrency network such as Bitcoin \cite{satoshibitcoin} or Ethereum \cite{wood2014ethereum} or a private network such as Ripple \cite{rippleoverview}  are examples of decentralized transaction network. The focus of this paper is on Blockchain based payment networks even though some of the results can be applied to other payment processing networks such as credit networks. Blockchain based payment networks have the potential of being a more secure and private way of transacting and promise significantly cheaper transaction costs by eliminating the premium paid to the central authority for each transaction. 
However, Scalability is a major challenge in such networks. Visa processes 1667 transactions per second \cite{bitcoinvsvisa}. On the other hand, Bitcoin, is limited to about 3-7 transactions per second \cite{bitcoinvsvisa} due to the underlying Proof of Work paradigm that secures Bitcoin. Moreover, Proof of Work involves computing cryptographic hashes, called mining which is not only expensive, but incurs huge energy costs.

Solutions such as Proof of Stake \cite{ppcoinstake} are proposed to address scalability issues in Blockchain \cite{scalingblockchain} and this is an area of active research. One way of addressing the scalability issues for the purpose of processing payments is to build a second payment processing networks on top of a Blockchain such as Bitcoin. Most of the transactions are processed in this second network, while the Blockchain with its expensive mining is used sparingly. Such a payment network is composed of bidirectional peer to peer payment channels, which are secured using smart contracts. Examples of such payment processing networks include  Lightning network for Bitcoin \cite{lightningnetwork}, Raiden network for Ethereum \cite{raiden}, and Atomic swap for inter Blockchain transactions \cite{atomicswap}. The focus of this paper is such Blockchain based secondary payment processing networks. Routing algorithms for such payment processing networks are studied in
Silent Whispers \cite{silentwhispers} and Speedy Murmurs \cite{speedy}.

The essential idea behind such networks is illustrated by the following example. Two agents, say Alice and Bob create a channel with capacity $\$c$ by each buying Bitcoins (or equivalently, recording a transaction in the underlying Blockchain) worth $\$c$. One may think of this channel as a kind of joint account worth $\$2c$ between Alice and Bob. Whenever, Alice wants to send Bob $\$a<\$c$, they add a new smart contract adjusting their individual ownership in the $\$2c$ Bitcoins, so that Alice owns $\$(c-a)$ and Bob owns $\$(c+a)$. The smart contract protects the transaction between Alice and Bob. Later, if Bob wants to send Alice some money, a new smart contract will be added reflecting effective ownership. However, due to the limit on the channel capacity, they can only send at most $\$c$ in each direction in a time slot. Each transaction takes some finite time, which defines the duration of a time slot. Moreover, the difference in total amount of money sent in the two directions can be at most $\$c$. If Alice and Bob each want to send the other $\$b>\$c$, they first send each other $\$c$ in one time slot and buffer the remaining amount. At the end of one time slot, the channel is balanced and the buffered requests can be sent  in the following time slots. In this approach, the underlying Bitcoin (Blockchain) network is used only to create a payment channel or to close a payment channel. All the other transactions are processed in the second overlay networks using smart contracts. 

Now suppose that, in addition, Bob and Chris have a channel between them, but Alice and Chris don't. Suppose Alice wants to send payments to Chris, Alice could send them to Bob, and Bob could forward them to Chris. Bob will be paid a premium to help process the transaction. In general, a transaction can be processed by finding a path between the sender and the receiver in the payment graph, where the agents are vertices and the channels are edges. All these transactions do not need to go through the underlying Blockchain, so this is known as off-chain rebalancing. However, note that this works only if on average, the requests in each direction on a link are balanced. Suppose there is a large amount of buffered requests from Alice to Bob, and no outstanding requests from Bob to Alice. These requests can only be met when there are corresponding future requests from Bob to Alice. One can then send some payment through the underlying Bitcoin network (Blockchain) to reduce the `imbalance' on this channel. Such a transaction is called on-chain rebalancing, and is usually more expensive because it involves cryptographic mining. 

There are several issues in the design of such payment networks, such as privacy, security, efficiency, etc \cite{silentwhispers} \cite{speedy} \cite{spider}. The focus of this paper is in designing a path from the sender to the receiver of each transaction so that the overall throughput of the network is maximized while limiting the use of on-chain rebalancing. While the routing problem is studied in the literature \cite{silentwhispers,speedy,spider}, a fluid model was used and a static scheduling policy is proposed. While static policies that do not use the system state while making decisions can maximize throughput, they are known to be sub-optimal in terms of other metrics such as queue lengths and delays. In this paper, we propose a stochastic model of the payment processing network that keeps track of the buffered requests and the imbalance on the channels. 

When a transaction is processed using off chain rebalancing on a channel, the individual ownership changes, and needs to be updated. Since we are interested in the long term throughput behaviour of the system, the system is sustainable only if on average, equal number of payments are made in both directions of each channel. Therefore, we make the simplifying assumption that at any time, on each channel, a request is served in one side, if and only if a request of equal value is served in the opposite direction. For example, if Alice wants to send $\$a$ to Bob and Bob wants to send $\$b$ to Alice and $\$c$ is the capacity of their channel, then, $\$\min\{a,b,c\}$ is served in both directions and the remaining request is buffered. The requests are buffered until there is enough demand in the opposite direction or there is enough capacity on the channel. This ensures that we don't have to constantly update the ownership of the channel, and makes the analysis simpler. The channels can thus be naturally modeled by a two-sided queue. 

Unlike traditional queues, where a server is fixed, and serves arriving demand, in a two-sided queue, both requests and servers arrive and are queued up. Each request is paired up with a server and both instantaneously depart from the system at the finish of service. Note that the `servers' in this context also correspond to requests in the payment network, since there is no conceptual distinction between the arrivals on the two sides of a two-sides queue. Even though there are various notions of stability, positive recurrence of the underlying Markov Chain is a strong form that ensures the existence of a steady-state distribution, and enables one to study the stationary queue lengths, delays, mean imbalance etc. 
Even if the arrival rates in both directions are equal, a two-sided queue is only null-recurrent and not positive recurrent. We need an external control to make it positive recurrent. Therefore, we use on-chain rebalancing to define stability based on positive recurrence in such a system.

\textbf{Main Contributions:} The main contributions of this paper are listed below.
\begin{enumerate}
    \item We propose a Stochastic Model to study routing of requests in a Blockchain based payment processing networks. Such a Stochastic Model enables the study of performance of state dependant or closed loop routing policies.
    \item We model such a payment processing network as a network of two-sided queues,
    and define stability of the network based on positive recurrence using the notion of on-chain rebalancing. We then characterize the capacity region of the payment processing network consisting of all demand rates under which the system can be made stable.
    \item We propose a novel state dependant policy and prove that it is throughput optimal, i.e., it stabilizes the system under any arrival rate in the capacity region. Since this is a stronger notion of stability than the rate stability studied in \cite{ganjali2005cell}, one can study steady-state behaviour of various metrics of interest.
    \item Under our proposed routing policy, we obtain an upper bound on the total buffered payment requests, and discuss its trade off with the amount of on chain rebalancing.
    \item We study the performance of the proposed algorithm using simulation on Ripple network data-set. The proposed algorithm routes $\sim 65\%$ of the incoming demand using only off-chain rebalancing and we note that it needs lesser amount of on-chain rebalancing than  Spider Network \cite{spider}.
\end{enumerate}
\textbf{Notation:} Now, we introduce some general notation used in this paper. We denote real numbers by $\mathbb{R}$ and non negative real numbers by $\mathbb{R}_+$. Integers are denoted by $\mathbb{Z}$, whereas the non negative integers are denoted by $\mathbb{Z}_+$. We denote $\max\{0,A\}$ by $[A]^+$ and indicator function denoted by $\mathbbm{1}_A$ is 1 if $A$ is true or 0 otherwise. For a collection of scalar variables $x$ indexed from some set, we use a bold $\mathbf{x}$ to denote a vector containing all its components. {Whenever a variable is implicitly dependent on $t$, we omit this dependence for the ease of notation.}

\section{Related Work} \label{sec:related work}
In this section, we present the prior work on payment processing networks, two-sided queues and MaxWeight family of algorithms. 

Lightning network is a peer-to-peer path based transaction network built on top of Bitcoin \cite{lightningnetwork}. It introduces a privacy preserving method to process transactions between two users via possibly routing it through multiple users. It however, does not provide a method to choose a path between a sender and receiver. Then, \cite{wu2016privacy} presents a privacy preserving algorithm to find a shortest path between two nodes in a centralized manner. Later, SilentWhispers \cite{silentwhispers} presents a privacy preserving protocol to route transactions in a decentralized manner using the idea of landmark routing based credit network design \cite{tsuchiya1988landmark}. In short, SilentWhispers is a routing, payment and an accountability algorithm in decentralized transactions routing to ensure privacy. It was the first distributed, privacy-preserving credit network. Speedy Murmurs \cite{speedy} is a more efficient routing algorithm which guarantees similar privacy levels. After that, \cite{spider} looked at routing cryptocurrency with the Spider network. Their simulation results show that Spider improves the volume and number of successful payments substantially comparing to the state-of-the-art.

Two-sided queues are studied in a variety of different contexts in the literature. Since the notion of stability in two-sided queues is not straight forward and one needs external control to achieve positive recurrence, several different approaches are taken in the literature. 
Spider network \cite{spider} which studied a fluid model of the payment processing network is the closest work in the literature. Rate stability was used there to bypass the issues about positive recurrence. One limitation of this approach is that rate stability does not ensure the existence of a steady-state distribution. 
A general framework of matching queues that subsumes two sided queues was studies in \cite{matchingqueues}. But the focus in \cite{matchingqueues} is to minimize the queue lengths over a finite time-horizon, and not longer term throughput. 

Two-sided queues naturally arise in ride-hailing systems, where riders are the requests and drivers are the servers. Closed queuing network models were used in \cite{banerjeeridehailing} and  \cite{yashkanoriaridehailing} to study ride-hailing systems when the total number of cars (servers) in the system are fixed. Issues about recurrence are avoided by dropping customer demand to make sure that the rider queues does not blow up to infinity. External control in terms of on-demand servers was used in   \cite{ondemandservers} to make the system stable.
Caldentey et al. \cite{caldentey2009fcfs} and Adan and Weiss \cite{adan2012exact} looked at a simplifying model which can be thought of as allowing customer arrivals only when there are servers waiting, and show positive recurrence under this model.
Two sided queues for Kidney exchanges were studied in \cite{roth2007kidney} \cite{yashkanoriabarter} and for barter exchanges with dynamic matching were studied in \cite{akbarpour2017thickness}. 

Tassiulas and Ephremides \cite{tassiulas1990stability} proposed the celebrated MaxWeight algorithm for scheduling in downlink in mobile base stations 
and its generalization, the Backpressure algorithm for routing and scheduling in multihop wireless networks \cite{tassiulas1990stability}. This then led to a huge line of literature on resource allocation problems in wired and wireless networks and the book \cite{srikantbook}, presents an excellent exposition. MaxWeight family of algorithms arise naturally when one views the scheduling problem as a fluid-like optimization problem, and consider a gradient descent algorithm on its dual. However, unlike open-loop policies that one obtains in the fluid limit, here, one obtains a closed loop or state dependant policies that are shown to maximize throughput using Foster-Lyapunov Theorem. In addition to networking problems, these ideas have been used in other resource allocation problems such as Cloud computing \cite{maguluri2016heavy}, Online ad matching \cite{srikantad}, ride sharing \cite{kanoria2019backpressure} etc.


\section{Model and Definitions  \label{sec:model}}
In this section, we first present the model and the definitions of stability. We will also define the capacity region of the system and show that no family of algorithm can make the system stable for the demand rate outside the capacity region.

\subsection{The Model}
We consider a payment processing network consisting of payment channels based on a Blockchain such as Bitcoin or other cryptocurrency systems. We model the payment processing network by a payment graph, $G(V,E)$, where, the vertices $V$  are the agents in the network and the edges $E$ are the payment channels between the agents. We assume that the graph $G$ is connected. All the edges in this graph are bidirectional. This means that we have a directed graph such that if $(u,v)\in E$, then, $(v,u)\in E$. 
Each edge $(u,v)\in E$ has a capacity $c_{(u,v)}$ such that $c_{(u,v)}=c_{(v,u)}$.  We consider this for the ease of exposition and our results can be easily extended for unequal capacities and unidirectional edges. Also define $c_{max}=\max_{(u,v) \in E} c_{(u,v)}$. We consider a discrete time model and let $a_{ij}(t)$ denote the value of payment request that arrives at time $t$ and should be sent from agent $i$ to agent $j$.
The demand $a_{ij}(t)$ is assumed to be in $\mathbb{Z}_+$ since it may be expressed in terms of smallest possible denomination such as cents, Satoshi for Bitcoin \cite{satoshiunit} and Wei for Ethereum \cite{wood2014ethereum}. The arrivals are assumed to be an iid sequence of random variables 
with mean $\lambda_{ij}$ and finite variance. Moreover, $a_{ij}(t)$ are assumed to be independent for different $i-j$ pairs. 

Each edge $(u,v)$ in the graph $G$ is a bidirectional payment channel between agents $u$ and $v$ established using a cryptocurrency system such as the Blockchain. Each edge $(u,v)$ is modeled as a two-sided queue consisting of two buffers of  outstanding demand from $u$ to $v$ and $v$ to $u$. The value of outstanding payment requests in these buffers is denoted by $q_{(u,v)}$ and $q_{(v,u)}$ respectively.

The focus of this paper is on designing a routing algorithm. For each payment request that arrives from $i$ to $j$, the goal is to find one or more paths from $i$ to $j$ in the graph $G$ on which this payment will be sent. We denote by $P_{ij}$ the set of all the possible paths from $i$ to $j$. Since we assume that the graph is connected, the set $P_{ij}$ is nonempty  for all $i\neq j \in V$. Each element in $P_{ij}$ is a set of non repeated edges which connects $i$ to $j$. Demand arising between $i$ and $j$ can be met using multiple paths in the set $P_{ij}$. We assume that the set of valid  paths $P_{ij}$ is given. This can be the set of all possible paths from $i$ to $j$ in the case of decentralized path based transaction networks such as \cite{spider}. In partially centralized networks such as \cite{silentwhispers}, the set $P_{ij}$ is restricted to the set of paths that pass through landmark nodes. 
Let $\mathcal{P}=\cup_{i\neq j \in V}P_{ij}$ denote the set of all valid paths.

{Now, we explain how the above model essentially represents an alternative state description of the real life model.}

{\textit{Real life model:} At the start of the time slot, each link in the transaction network has a capacity $c_{(u,v)}$ for all $(u,v) \in E$. In a given time slot, there is a transaction request $a_{ij}$ between agent $i$ and $j$. Now, the transaction request is routed using a path from $i$ to $j$ which has enough capacity. Then, the capacity of each link is updated accordingly. The length of a time slot is chosen based on the delay incurred to process the transaction. For example, consider the capacity of all the edges to be \$5. Now, Alice sends \$1 to Charlie via Bob which will update the capacity of Alice-Bob and Bob-Charlie link to \$4 in one direction and \$6 in another and the transaction will be processed in one time slot. If the transaction is in excess of the capacity, it is either dropped, or buffered to be completed in the future time slots.}

{\textit{In our model}, we consider all the transactions that are routed through a path is buffered in the queues at the edges in the path. The transaction are processed only when transaction requests are routed through the same edge in the opposite direction. Thus, in the above example, rather than routing the transaction and changing the capacities of the link, we add \$1 to the Alice-Bob and Bob-Charlie queue length and the capacity of each link remains to be \$5 on both the directions. This is just an alternative way of describing the state of the system. In particular, given $q_{(u,v)}$, $q_{(v,u)}$ and $c_{(u,v)}$, it can also be interpreted as the capacity of edge $(u,v)$ being $[c_{(u,v)}-q_{(u,v)}]^+$ and the excess transaction waiting in the buffer being $[q_{(u,v)}-c_{(u,v)}]^+$. With the interpretation we consider, the service in both directions becomes equal as the transactions is completed only when there are transactions waiting to be processed in both the directions which justifies \eqref{eq:suv}. This is equivalent to the real life model since we are interested in the long term behaviour. }

The routing algorithm determines the amount of payment requests $x_{p}(t)$ to be routed through each  path $p \in P_{ij}$. Note that the assignments $x_{p}(t)$ should be picked such that $\sum_{p\in P_{ij}} x_{p}(t) = a_{ij}(t)$ for all $i,j \in V$, so that all the arriving requests are assigned to some path.
Once the amount of payments $x_{p}(t)$ are determined for each path $p \in \mathcal{P}$ , a request of  $x_{p}(t)$ is added to all the buffers on the path $p$. 
We use $y_{(u,v)}(t)$ to denote the total amount of payment request that is added on the channel $(u,v)$ at time $t$, i.e., 
\begin{align}
    y_{(u,v)}(t)={}&\sum_{p \in \mathcal{P} : (u,v)\in p} x_p(t) \label{eq: yuv}.
\end{align}

At each time $t$, on each edge $(u,v)$, the amount of service is the maximum possible values up to the channel capacity $c_{(u,v)}=c_{(v,u)}$, such that an equal amount of requests on either side of each channel are served. If $s_{(u,v)}(t)$ denotes the amount of payment requests served in buffer $q_{(u,v)}$, then, we have 
\begin{align}
 s_{(u,v)}(t)={}&\min\bigg\{q_{(u,v)}(t), q_{(v,u)}(t),c_{(u,v)}\bigg\} . \label{eq:suv}
\end{align}
Note that by the symmetry in the definition, we have that $s_{(u,v)}(t)=s_{(v,u)}(t)$.

In addition to routing, a second control decision that needs to be made is about on-chain rebalancing. Recall that most of the requests are served through smart contracts in the payment processing networks, and bypass the use of the underlying Blockchain and is called off-chain rebalancing. However, some of the demand may sometimes be served by going through the Blockchain, and this is called on-chain rebalancing, and is more expensive. The second control decision is to decide how much of the outstanding demand in each direction on each channel is met using on-chain rebalancing. The goal is to minimize the use of on-chain rebalancing. Let $r_{(u,v)}(t)$ denote the amount of on-chain rebalancing on the link $(u,v)$ at time $t$. The outstanding demand in each buffer evolves as
\begin{align}
    q_{(u,v)}(t+1)={}&q_{(u,v)}(t)+y_{(u,v)}(t)-s_{(u,v)}(t)-r_{(u,v)}(t). \label{qu}   
\end{align}
We assume that the amount of on-chain rebalancing satisfies $r_{u,v}(t)\leq q_{(u,v)}(t)+y_{(u,v)}(t)-s_{(u,v)}(t)$ so that the outstanding requests, $q_{(u,v)}(t+1)$ do not go negative. 
Even though we say routing algorithm, we consider algorithms that perform routing as well as on-chain rebalancing.
For a given routing algorithm, we say that the rate of on-chain rebalancing is at most $\epsilon$ if the long term average amount of on-chain rebalancing on all links is smaller than $\epsilon$ with probability 1, i.e., $\limsup_{T\to \infty}\left(\sum_{t=1}^T \sum_{(u,v)}r_{u,v}(t)\right)/T \leq \epsilon $ w.p. 1.


\subsection{Stability and Capacity Region}

Consider a classical queue, where server serves requests at rate $\mu$ and the requests arrive according to some stochastic process at rate $\lambda$. As long as the arrival rate is smaller than the service rate, i.e., $\lambda <\mu$ the system is usually stable, for various different notions of stability. 

However, the situation in a two-sided queue is more involved. We now have two queues, and requests in both the queues are paired for service. As an example, consider a payment network consisting of a single channel $(u,v)$ with infinite capacity, and iid Bernoulli payment requests on both sides. This is a Discrete Time Markov Chain (DTMC). If the rate of arrivals on each side is different,  i.e., if $\lambda _{(u,v)}>\lambda _{(v,u)}$ or $\lambda _{(u,v)}<\lambda _{(v,u)}$ , then the system is transient and so not stable. When $\lambda _{(u,v)}=\lambda _{(v,u)}$, the two-sided queue is said to be rate-stable. However, the underlying DTMC is still not very well-behaved, because it is null-recurrent since it is a symmetric random walk in one dimension. In particular, there is no steady-state distribution, and the limiting buffer lengths may not be bounded, i.e, $\lim_{C \to \infty} \lim_{t \to \infty} P(q_{(u,v)}(t)+q_{(v,u)}(t) \geq C) \neq 0$. Null-recurrence is not a consistent notion of stability. This is because of the following: now consider a network consisting of three channels similar to $(u,v)$ that are not connected to each other, and so are  independent. This system is equivalent to a symmetric random walk in three dimensions, which is transient. Therefore, we cannot take null-recurrence as a notion of stability.

In this paper, we will consider two notions of stability, that are both stronger than rate stability that are studied in literature \cite{ratestability}. 
One is in terms of positive recurrence of the DTMC, and the other is that $\lim_{C \to \infty} \lim_{t \to \infty} P(q_{(u,v)}(t)+q_{(v,u)}(t) \geq C) = 0$. 
The previous example shows that without external control, a two-sided queue is not stable under either of these notions. In other words, off-chain rebalancing alone cannot make the payment network stable. So, we need at least some on-chain rebalancing to make the system stable. However, it turns out that an arbitrarily small amount of on-chain rebalancing is enough. Therefore, instead of studying routing algorithms, we study families of algorithms, where for any given $\epsilon>0$, each family consists an algorithm with on-chain rebalancing rate smaller than $\epsilon$.
We first present the weaker notion of stability.

\begin{definition}[Weak Stability]
Consider the payment processing network with a given arrival rate vector $\vlam$. A family of routing algorithms is said to weakly stabilize the network if for any $\epsilon>0$, there exists an algorithm in the family that uses at most  $\epsilon$ rate of on-chain rebalancing such that, $\lim_{D \to \infty} \lim_{t \to \infty} \mathbb{P}(\sum_{(u,v)\in E}q_{(u,v)}(t) \geq D) = 0$ .
\end{definition}

In classical single sided queuing systems, weak stability is defined as the set of arrival rate vectors for which there exists an algorithm under which we have $\lim_{D \to \infty} \lim_{t \to \infty} \mathbb{P}(\sum_{(u,v)\in E}q_{(u,v)}(t) \geq D) = 0$. Although, in our case we need to consider a family of algorithms to define weak stability due to the following two reasons:
\begin{itemize}
    \item With zero on chain rebalancing, it is not possible to stabilize the system as discussed above.
    \item With arbitrary on chain rebalancing, it is meaningless to talk about stabilizing the system as after routing incoming demands using any algorithm, the excess demand can be routed using on-chain rebalancing and the system can be stabilized under \textit{any} algorithm.
\end{itemize}
Thus, we need a \textit{family} of algorithm such that the system is stabilized under any given constraint on on-chain rebalancing. Based on our notion of stability, we now define the capacity region. 
\begin{definition}[Capacity Region]
The capacity region for the payment processing networks is the set of arrival rate vectors $\vlam$ that are weakly stabilizable by some family of routing algorithms.
\end{definition}

We are interested in families of algorithms that stabilize the network for the maximum possible set of arrival rates, and so we now define throughput optimality. 
\begin{definition}[Throughput Optimality]
A family of algorithms is said to be throughput optimal if it can weakly stabilize the payment processing network under any arrival rate in the capacity region. 
\end{definition}



Let $\mathcal{C}$ define the set of arrival rates $\vlam$ if there  exists an $\mathbf{x} \in \mathbb{R}_{+}^{|\mathcal{P}|}$ such that,
\begin{align}
     \sum\limits_{p \in P_{ij}} x_p &= \lambda_{ij} \ \forall i\neq j \in V,  \label{prop1}\\
    \sum\limits_{p \in P:(u,v) \in p} x_p+\sum\limits_{p \in P:(v,u) \in p} x_p &\leq 2c_{(u,v)} \ \forall (u,v) \in E, \label{prop2} \\
 \sum\limits_{p \in P:(u,v) \in p} x_p-\sum\limits_{p \in P:(v,u) \in p} x_p &=  0  \ \forall (u,v) \in E.  \label{prop3}
\end{align}

This essentially says that the set $\mathcal{C}$ consists of all the demands rates $\vlam$, such that they can be allocated to various paths, so that the resulting request rates on each channel are balanced in both directions and respect the capacity constraints of the channel. This set of inequalities is analogous to the capacity region defined in \cite{spider}.  

To get an intuition for the definition of $\mathcal{C}$, let $\mathcal{C}^{\infty}$ denote the set $\mathcal{C}$ when all the capacities are infinity. It is easy to see that 
\begin{align}
\mathcal{C}^{\infty}=\bigg\{\Lambda: \sum_j \lambda_{ij} = \sum_j \lambda_{ji} \ \forall i \in V \bigg\}. \label{Cinfty}
\end{align}

This just says that the demand rates should be such that in the long run, the total incoming demand to a vertex should be equal to the total outgoing demand or equivalently, we can say that, the demand has to be a circulation \cite [Prop. 1]{spider}.

We will see that $\mathcal{C}$ is indeed the capacity region of the payment processing network. First we present the following proposition to show that $\mathcal{C}$ is contained in the capacity region.

\begin{proposition}\label{prop}
Any rate vector $\vlam \notin \mathcal{C}$ is not weakly stabilizable by any family of routing algrithms. 
\end{proposition}

The proof of the proposition is based on strong law of large numbers, and is similar to the argument used in other systems, such as Theorem 4.2.1 in \cite{srikantbook}, 
and is deferred to Appendix \ref{app:proofofprop}.

We show that $ \mathcal{C}$ is indeed the capacity region in Theorem \ref{theorem} in the next section, by presenting a family of routing algorithms that stabilize the payment network for any $\vlam \in \mathcal{C}$. However, we will show that this family exhibits a stronger form of stability defined as follows.

\begin{definition}[Strong Stability]
Consider the payment processing network with a given arrival rate vector $\vlam$ and a family of algorithms under which the queue lengths vector $\mathbf{q}$ is a DTMC. 
This family is said to strongly stabilize the network if for any $\epsilon>0$, there exists an algorithm in the family that uses at most  $\epsilon$ rate of on-chain balancing, and the DTMC is positive recurrent. 
\end{definition}

Note that strong stability implies weak stability because positive recurrence implies existence of a stationary distribution. Therefore, once it is established in Theorem \ref{theorem} that $\vlam \in \mathcal{C}$ is strongly stabilizable and so weakly stabilizable, we have that both the notions of stability gives the same capacity region $\mathcal{C}$.

\section{Throughput Optimal Routing \label{sec:throughput optimality}}

In this section, we first present the proposed algorithm and then prove that it is throughput optimal using the Foster-Lyapunov Theorem.


\subsection{Routing Algorithm}

Before we present the algorithm, we first present one more definition. 
We define the imbalance, $z_{(u,v)}=q_{(u,v)}-q_{(v,u)}$ as the difference in the outstanding demand from $u$ to $v$ and $v$ to $u$ on each edge $(u,v)$ in the graph. It denotes how much more outstanding requests are from $u$ to $v$ than from $v$ to $u$. Note that the imbalances can be positive or negative and in particular, $z_{(v,u)} = -z_{(u,v)}$.


The evolution of the DTMC $\mathbf{q}(t)$ is completely characterized by \eqref{qu}, once the routing policy gives the path allocations $\mathbf{x}(t)$ and the amount of on-chain rebalancing $\mathbf{r}(t)$. The proposed family of routing algorithms with parameters $M$, $\delta$ and $\alpha$ is given in Algorithm \ref{alg2}.


\begin{algorithm}
\caption{Routing Algorithm}\label{alg2}
\begin{algorithmic}
\STATE \textbf{Parameters:} $M>c_{max},\delta \leq 1$, $\alpha>1$
\STATE \textbf{Input:} $\mathbf{q}(t), \mathbf{a}(t)$ 
\STATE \textbf{Initialize:} $\mathbf{x}(t)=\mathbf{0},\mathbf{r}(t)=\mathbf{0}$
\STATE \COMMENT{Routing of the demand}
    \FOR{$i \neq j \in V$: $a_{ij}(t)>0$} 
        \STATE 
        $p^{*}=\arg\min_{p \in P_{ij}}\big\{\sum_{(u,v) \in p}   \big(z_{(u,v)}(t)$ 
        \STATE \hspace{2cm}$ +\coeff{2}( q_{(u,v)}(t) 
            + q_{(v,u)}(t) )\big)\big\}$ \\
     $x_{p^{*}}(t)=a_{ij}(t)$ 
    \ENDFOR
\STATE \COMMENT{On-chain Rebalancing}
    \FOR{$(u,v) \in E$}
        \IF{$q_{(u,v)}(t)>M$} 
            \STATE $r_{(u,v)}(t)=\begin{cases}
           1 & \textit{w.p. } \delta \\
           0 & \textit{w.p. } 1-\delta \end{cases}$
        \ENDIF
    \ENDFOR
\STATE \textbf{Output:} $\mathbf{x}(t),\mathbf{r}(t)$
\end{algorithmic}
\end{algorithm}

For each source destination pair with an arrival, the algorithm finds a path $p \in P_{ij}$ that has the minimum cost, where the cost of each edge $(u,v)$ is taken to be $z_{(u,v)}(t)+\coeff{2}(q_{(u,v)}(t)+q_{(v,u)}(t))$. Thus, this algorithm falls into the class of MaxWeight algorithms, and can be intuitively thought of as follows. Since the requests on each of our channels are served by pairing them in both the directions, the algorithm tries to equalize the outstanding requests on both sides of the channels by adding more requests to the less loaded side of the channels that have the greatest imbalance.  Moreover, due to the capacity constraints, one needs to make sure that the load is distributed appropriately along various paths. So, the algorithm gives more priority to less loaded paths, by adding $\coeff{2}(q_{(u,v)}(t)+q_{(v,u)}(t))$ to the weight of each link. Here the coefficient of the weight of the load of the path compared to the imbalance can be tuned using the parameter $\alpha$. For a given system, an appropriate $\alpha$ can be chosen by conducting simulations on the system. Note that the proposed algorithm picks a single path for each source destination pair. The results can be extended to develop algorithms that pick several paths for privacy reasons, and is a future research direction.

Then the algorithm assigns 1 unit of on chain rebalancing with probability $\delta$ for all links $(u,v)$ which are such that $q_{(u,v)}(t)>M$. Note that, the only technical condition we need is the expected on chain rebalancing is $\delta$ if the queue length is greater than a certain threshold $M$ which can be realized by performing randomized on chain rebalancing using any distribution with the expected value of the distribution equal to $\delta$. Here, $M$ and $\delta$ are parameters for the family of algorithms. Given $\epsilon>0$, we can tune $M$ and $\delta$ so that the on chain rebalancing rate is at most $\epsilon$.

Finally, note that the algorithm can be easily generalized to have different thresholds $M$ for different edges $(u,v) \in E$ with the condition that $M_{(u,v)} \geq c_{(u,v)}$. Although, we consider a single threshold $M$ for the ease of exposition.

\subsection{Throughput Optimality}
In this subsection, we will prove the main theorem of our paper that the family of algorithms in Algorithm \ref{alg2} are throughput optimal, and that $\mathcal{C}$ is indeed the capacity region of the payment processing network.

To ensure the DTMC is irreducible, we restrict our attention to the set of states $\mathcal{S}$ which can be reached from the state $\mathbf{q}= \boldsymbol{0}$ under the Algorithm 1 and make the following assumption on the arrivals. 

\begin{assumption} \label{assumption}
The arrivals  $\mathbf{a}$ satisfy
\begin{align}
    \mathbb{P}[a_{ij}=0]&>0 \ \forall i\neq j \in V, \label{ass1} \\
    \mathbb{P}[a_{ij}=1]&>0 \ \forall i\neq j \in V. \label{ass2}
\end{align}
\end{assumption}

Equation \eqref{ass1} ensures that there is a non zero probability that from any state, $\vq_{(u,v)}$ will come back to zero and \eqref{ass2} ensures that there is a non zero probability that the arrivals are such that $q_{(u,v)}$ becomes equal to $q_{(v,u)}$ for all $(u,v) \in E$. Once this happens, there is a non zero probability that $q_{(u,v)}$ will come back to zero by \eqref{ass1}. So, the DTMC becomes irreducible on $\mathcal{S}$ under \eqref{ass1} and \eqref{ass2}. We now present the main Theorem of the paper. 
\begin{theorem}
\label{theorem}
Consider the payment processing network with arrival rates $\vlam \in \mathcal{C}$, operating under Algorithm \ref{alg2} with parameters $M>c_{max}$, $1\geq \delta>0$ and $\alpha>1$ such that $\delta<M$.
Then the DTMC $\{\mathbf{q}(t): t \in \mathbb{Z}_+\}$ is positive recurrent with an on-chain rebalancing rate of at most $O(\delta)$. Thus the family of algorithms in Algorithm \ref{alg2} is strongly stable. Moreover, under Algorithm \ref{alg2}, in steady state, we have

\begin{align}
    \sum_{(u,v) \in E}&\mathbb{E}[q_{(u,v)}] \leq \hspace{-1.3pt} \bigg(12|E|\hspace{-3pt}\sum_{i \neq j \in V}\sigma_{ij}^2+2\hspace{-3pt}\sum_{(u,v) \in E} c_{(u,v)}^2\bigg)\frac{\alpha c_{max}}{\delta^2} \nonumber \\*
    & \hspace{-1.3pt}+12|E|\frac{\alpha c_{max}}{\delta}+(4\alpha c_{max}|E|)\frac{M}{\delta}+\frac{c_{max}|E|}{c_{min}}M. \label{eq: expectedqueuelength}
\end{align}

\end{theorem}
Note that the strong stability of the family of algorithms in Algorithm \ref{alg2} in the theorem follows by making the on-chain rebalancing rate arbitrarily small by choosing small enough $\delta$. Together with Proposition \ref{prop}, the Theorem establishes that $\mathcal{C}$ is the capacity region of the payment processing network. Thus the family of algorithms in Algorithm \ref{alg2} is throughput optimal. Also, note the trade off between the rate of on chain rebalancing used which is $O(\delta)$ and the expected sum of queue lengths with respect to the steady state distribution of the DTMC which is $O(\frac{1}{\delta^2})$. So, with decreasing $\delta$, we reduce the use of on chain rebalancing but increase the time it takes for the transactions to process.

The proof of the theorem is deferred to appendix. Although, we present the sketch of the proof in the main body of the paper. Before that, we state Lemma 1 which will be useful for the proof. 

The following lemma connects the min cost path, the arrival rates with the capacities and the queue lengths at the links, as long as the $\vlam$ is in the capacity region $\mathcal{C}$. In fact Lemma 1 encapsulates all the properties of the capacity region $\mathcal{C}$ that we need for the proof of Theorem \ref{theorem}.

\begin{lemma} \label{lemma}
If the demand matrix is in the capacity region, i.e. $\vlam \in \mathcal{C}$, for any $\mathbf{q}$, we have
\begin{align*}
      \sum_{i \neq j \in V} \hspace{-0.25 em} \lambda_{ij}\min_{p \in P_{ij}}&\bigg\{ \hspace{-0.25 em} \sum_{(u,v) \in p} \big(z_{(u,v)}+\coeff{2}(q_{(u,v)}+q_{(v,u)})\big)\bigg\} \nonumber \\
      &\leq \sum_{(u,v) \in E} \frac{\delta q_{(u,v)}}{\alpha} . \nonumber
\end{align*}
\end{lemma}
The proof uses basic duality theory and the reader can refer to the appendix for the details. We only present the sketch of the proof here.
\begin{proof}[Sketch of the Proof]
The proof is carried out in the following steps:
\begin{itemize}
    \item We start by defining a linear program with objective function to be zero and the constraints to the constraints of the capacity region. 
    \item Then, as $\vlam \in \mathcal{C} $, the primal is feasible and we also argue that the dual is feasible by presenting a feasible point for the dual.
    \item Then we use weak duality which gives us the dual objective function to be non-negative for all the feasible solutions of the dual LP.
    \item Then we carefully pick feasible dual variables which gives us the Lemma.
\end{itemize}
\end{proof}
{Intuitively, by carrying out $\delta$ on-chain rebalancing, we are introducing a drift towards zero. Now, if $\delta$ is small, the introduced drift towards zero is small which will lead to higher queue lengths. This is reflected by Theorem \ref{theorem} as the expected sum of queue lengths is $O(1/\delta^2)$. The trade off between the drift and expected sum of queue length in queueing theory is fundamental. For example, for a single server queue, by Kingman's bound, the expected queue length is $O(1/\epsilon)$ for drift (difference between service rate and arrival rate) equal to $\epsilon$.} Now, we present the sketch of proof of Theorem \ref{theorem}.
\begin{proof}[Sketch of the Proof of Theorem \ref{theorem}]
First we show that the DTMC $\{\mathbf{q}(t): t \in \mathbb{Z}_{+}\}$ is positive recurrent using Foster-Lyapunov Theorem. This can be done in the following steps:
\begin{itemize}
    \item First we carefully choose the Lyapunov function as
    \begin{align*}
        V(\xs)&\overset{\Delta}{=}\sum_{(u,v) \in E} z_{(u,v)}^2+\sum_{(u,v) \in E} \coeff{2}(q_{(u,v)}+q_{(v,u)})^2. 
    \end{align*}
    \item Next, we calculate the one-step drift ``$V(q(t+1))-V(q(t))$'' of the defined Lyapunov Function and bound the quadratic terms by known parameters.
    \item Then, we simplify the expression of the drift using Algorithm \ref{alg2} and Lemma \ref{lemma}.
    \item Then, we define a finite set of queue lengths and show that the one-step drift is negative outside this set. Thus, by Foster-Lyapunov Theorem, the DTMC is positive recurrent.
    \item Finally, we argue that the on-chain rebalancing rate is $O(\delta)$ which shows that the Algorithm \ref{alg2} is Throughput Optimal.
\end{itemize}
After showing positive recurrence, we use the moment bound theorem \cite{HajekComm} to bound the sum of expected queue lengths. This completes the proof.
\end{proof}

{While the core idea is similar to a typical MaxWeight proof, there are several differences here. In addition to showing the negative drift, we also characterize the amount of external control (on-chain rebalancing). In order to show negative drift, we had to use two different Lyapunov functions. In addition to the typical Lyapunov function that is quadratic in the queue lengths, we also use a quadratic function of the \textit{imbalance} ($z_{(u,v)}$) on each link. }
\section{The Fluid Analysis \label{sec:fluid}}
In this section, we want to provide the reader some intuition and a better understanding of Algorithm \ref{alg2} and present the framework which motivates Algorithm \ref{alg2}. Particularly, we will first present the fluid model and then solve it using dual descent technique. Finally we will draw similarities in optimizing the fluid objective and Algorithm \ref{alg2}.  
\subsection{The Fluid Model}
We will now analyze the system by neglecting all the variability.  For this, we will formulate the deterministic optimization problem which will find the optimal routes based on the transaction rates $\vlam$ between the vertices. The objective is to solve the fluid model and infer a stochastic algorithm using it. The Linear Program to find the optimal routes is,  \cite{spider}
\begin{equation} 
    \max \sum\limits_{i,j \in V} \sum \limits_{p \in P_{ij}} x_p \label{obj}
\end{equation}
subject to
\begin{align} 
    \sum_{p \in P_{ij}} x_p & \leq  \lambda_{ij} \ \forall i \neq j \in V \label{xp1} \\
    \sum_{p \in P:(u,v) \in p} x_p+\sum\limits_{p \in P:(v,u) \in p} x_p &\leq  2c_{(u,v)} \ \forall (u,v) \in E \label{capacity} \\
    \sum_{p \in P:(u,v) \in p} x_p-\sum\limits_{p \in P:(v,u) \in p} x_p &=  0 \ \forall (u,v) \in E \label{bal} \\
     x_p &\geq 0 \ \forall p \in \mathcal{P}.  \label{xp2}
\end{align}
Equation \eqref{obj} maximizes the throughput, i.e. the total amount of transaction routed using the off-chain rebalancing, where $x_p$ is the amount of transaction routed using the path $p$. Equation \eqref{xp1} enforces the condition that the total amount of transaction routed from sender $i$ to receiver $j$ is less than or equal to the transaction rate $\lambda_{ij}$. Equation \eqref{capacity} restricts the total amount of transaction that can be routed via any edge $(u,v) \in E$ by its capacity $2c_{(u,v)}$. Equation \eqref{bal} enforces the flow balance requirement on each edge in the network. 

To solve this Linear Program, we will write the Lagrangian of the problem. Let, $\mu_{(u,v)}$ to be the Lagrangian multipliers for \eqref{bal} and $\delta_{(u,v)}$ for \eqref{capacity}, the optimization problem becomes,

\begin{align}
    \max \sum_{i\neq j \in V} \sum_{p \in P_{ij}} x_p -  \sum_{(u,v) \in E}\mu_{(u,v)}  \big[ \sum_{p \in P: (u,v) \in p} x_p \nonumber\\ 
    - \sum_{p \in P: (v,u) \in p} x_p\big]-\sum_{(u,v) \in E}\delta_{(u,v)}  \big[ \sum_{p \in P: (u,v) \in p} x_p\nonumber\\
    + \sum_{p \in P: (v,u) \in p} x_p-2c_{(u,v)}\big]
\end{align}
subject to (\ref{xp1}) and (\ref{xp2}).
Simplifying the objective yields,
\begin{align}
    \max &\sum_{i\neq j \in V} \sum_{p \in P_{ij}} x_p\sum_{(u,v)\in p}\big(1+\mu_{(u,v)}-\mu_{(v,u)} \nonumber \\
    &-\delta_{(u,v)}-\delta_{(v,u)}\big)+2\sum_{(u,v) \in E} \delta_{(u,v)}c_{(u,v)}. \label{lag}
\end{align}
subject to (\ref{xp1}) and (\ref{xp2}).

The decision variables in \eqref{lag} is the amount of transaction $x_p$ to be routed through path $p$ for all paths $p \in \mathcal{P}$. We define the following variable for ease of notation:
\begin{align}
    w_p = 1+\sum_{(u,v) \in p}(-\mu_{(u,v)}+\mu_{(v,u)}-\delta_{(u,v)}-\delta_{(v,u)}) \ \forall p \in \mathcal{P}. \nonumber
\end{align}
Now, we will state the Algorithm 2 below, which takes the demand rate vector $\boldsymbol{\lambda}$ as the input and outputs the optimum vector $\mathbf{x}_p$ for all $p \in \mathcal{P}$ which maximizes the Lagrangian function \eqref{lag} subject to (\ref{xp1}) and (\ref{xp2}).

\begin{algorithm}
\caption{Maximizing the fluid objective}\label{alg1}
\begin{algorithmic}[2]
\STATE \textbf{Input:} $\boldsymbol{\lambda}$
\STATE \textbf{Initialize:} $\mathbf{x}=\mathbf{0}$
\FOR{$(i,j): \lambda_{ij} > 0$}
    \IF{$\max_{p \in P_{ij}} w_p > 0$}
        \STATE $p^*=\arg\max_{p \in P_{ij}} w_p$
        \STATE $x_{p^*} = \lambda_{ij}$ \COMMENT{Assigning to the upper bound}
    \ELSE
        \STATE $\sum_{p \in P_{ij}} x_p = 0$ \COMMENT{Assigning to the lower bound}
    \ENDIF
\ENDFOR
\STATE \textbf{Output:} $\mathbf{x}$
\end{algorithmic}
\end{algorithm}
If the value of dual variables (Lagrangian Multipliers: $\mu_{(u,v)}$ and $\delta_{(u,v)}$) are known, then we can apply the Algorithm \ref{alg1} and find the routes to maximize the fluid objective. To find the Lagrangian multipliers, we will write the dual of the Linear Program which involves the dual variables $\mu_{(u,v)}$ and $\delta_{(u,v)}$. The dual problem is,
\begin{align}
    \sum_{i\neq j \in V} \max_{p \in P_{ij}}\bigg[\sum_{(u,v) \in p}(1-\mu_{(u,v)}+\mu_{(v,u)}-\delta_{(u,v)} \nonumber \\
    -\delta_{(v,u)})\bigg]^+ \lambda_{ij}+2\sum_{(u,v) \in E} \delta_{(u,v)}c_{(u,v)} \label{dual} \\
    \textit{subject to} \hspace{0.35in} \mu_{(u,v)} \in \mathbb{R}, \delta_{(u,v)} \geq 0 \hspace{0.15in} \forall (u,v) \in E
\end{align}
Now, the dual descent technique is employed to find the dual variables by the following equations:
\begin{align}
    \mu_{(a,b)}(t+1)&=\mu_{(a,b)}(t)-\frac{1}{M_1}\frac{\partial D}{\partial \mu_{(a,b)}} \label{dualmu} \\
    \delta_{(a,b)}(t+1)&=\bigg[\delta_{(a,b)}(t)-\frac{1}{M_2}\frac{\partial D}{\partial \delta_{(a,b)}}\bigg]^{+}, \label{dualdelta}
\end{align}
    where D is the dual objective function \eqref{dual} and $\frac{1}{M_1}$ and $\frac{1}{M_2}$ are the step sizes of the iterations respectively. We used projected dual descent for $\delta_{(u,v)}$ due to the constraint $\delta_{(u,v)}\geq 0$. Solving \eqref{dualmu} and \eqref{dualdelta} we get,
\begin{align}
    \mu_{(a,b)}(t+1)={}&\mu_{(a,b)}(t)+\frac{1}{M_1}\sum_{i\neq j \in V}\lambda_{ij}\mathbbm{1}_{(a,b) \in p^*_{ij}} \nonumber
    \\ &- \frac{1}{M_1}\sum_{i\neq j \in V}\lambda_{ij}\mathbbm{1}_{(b,a) \in p^*_{ij}}, \label{queue} \\
     \delta_{(a,b)}(t+1)={}&\bigg[\delta_{(a,b)}(t)+\frac{1}{M_2}\sum_{i\neq j \in V}\lambda_{ij}\mathbbm{1}_{(a,b) \in p^*_{ij}} \nonumber
    \\ &+ \frac{1}{M_2}\sum_{i\neq j \in V}\lambda_{ij}\mathbbm{1}_{(b,a) \in p^*_{ij}}-2\frac{c_{(u,v)}}{M_2}\bigg]^{+}. \label{vqueue}
\end{align}
Here, $p^{*}_{ij}$ is the optimal route chosen using Algorithm 2 by which all the demand arising between $i$ and $j$ is routed. If for all $p \in P_{ij}$, the weight $\sum_{(u,v) \in p}(1-\mu_{(u,v)}+\mu_{(v,u)}-\delta_{(u,v)}-\delta_{(v,u)})$ is negative, then $p_{ij}^{*}=\phi$.

The behavior of $M_1\mu_{(a,b)}$ in \eqref{queue} has similar structure to the imbalance $z_{(a,b)}$ given in the proof of Theorem \ref{theorem} in the appendix. The difference is that $\lambda_{ij}$ is substituted by $\sum_{p \in P_{ij}} x_p(t)$, which is the exact amount of transaction that is met at time $t$. Thus, we will approximate $M_1\mu_{(a,b)}$ as $z_{(a,b)}$ on the edge $(a,b) \in E$. Also, the behaviour of $M_2 \delta_{(a,b)}$ has a similar structure to $q_{(a,b)}+q_{(b,a)}$ as in \eqref{qu}. Although, demand rate $\lambda_{ij}$ is replaced by actual demand and the service does not solely depend on the capacity of the link but also depends on the outstanding buffer in the opposite direction. So, $c_{(a,b)}$ is replaced by $s_{(a,b)}$. Also note that on chain rebalancing terms are not present in \eqref{queue} and \eqref{vqueue} as the Algorithm 2 only uses off chain rebalancing and rejects all the excess demand.

Optimal path in Algorithm \ref{alg2} can be obtained from Algorithm \ref{alg1} by appropriately substituting $M_1\mu_{(a,b)}=z_{(a,b)}$ and $M_2\delta_{(a,b)}=q_{(a,b)}+q_{(b,a)}$, ignoring the constants and scaling the term $M_2 \delta$ appropriately. Condition when $w_p \leq 0$ in Algorithm \ref{alg1} has been weakened by meeting a fraction of demand using on chain rebalancing rather than not meeting the demand ($x_p=0)$.

Note that, the inequality \eqref{capacity} in the primal corresponds to the traditional queue ``$q_{(u,v)}+q_{(v,u)}$'', where the inequality constraint can be thought of as the capacity constraint of the single queue that the arrival rate should be ``less than'' the service rate. 

Whereas, the equality constraint \eqref{bal} in the primal corresponds to the two sided queue ``$z_{(u,v)}$''. Here, the equality constraint can be thought of as the capacity constraint of the single two sided queue that the arrival rates in both the directions should be equal. It is also consistent with the fact that the dual variable of the equality constraint can be both positive and negative. Also, an equality constraint can be thought of as two inequality constraint which is analogous to a two sided queue equivalent to the difference of two single server queues.

\section{Modified Water-Filling Routing Algorithm} \label{sec: modalg}
In this section, we will further discuss Algorithm \ref{alg2} and present a variant of the algorithm.
It is desired that the routing algorithm satisfy the Payment Channel Network (PCN) protocol. The conditions being
\begin{itemize}
    \item Atomicity of the payments, i.e. the payments should be processed immediately or they are rejected.
    \item The funds in all the edges which are used to process the payment is locked until the whole transaction is processed.
\end{itemize}
Moreover, it is not desirable to carry out on-chain rebalancing often as it is slow compared to off-chain rebalancing.

We present a modification of the Algorithm \ref{alg2} which takes into account the PCN protocols and is applicable in real life systems.

\begin{algorithm}
\caption{Modified Water-Filling Routing Algorithm}
\label{alg: water_filling}
\begin{algorithmic}
\STATE \textbf{Parameters:} $M_{(u,v)}=c_{(u,v)} \ \forall (u,v) \in E,\delta \leq 1$, $\alpha>1$
\STATE \textbf{Input:} $\mathbf{q}(t), \mathbf{a}(t)$ 
\STATE \textbf{Initialize:} $\mathbf{x}(t)=\mathbf{0},\mathbf{r}(t)=\mathbf{0}, \mathbf{x}'(t)=\mathbf{0}$
\STATE \COMMENT{Routing of the demand}
    \FOR{$i \neq j \in V$} 
    \WHILE{$a_{ij}(t)>0$}
        \STATE 
        $p^{*}=\arg\min_{p \in P_{ij}}\big\{\sum_{(u,v) \in p}   \big(z_{(u,v)}(t)$ 
        \STATE \hspace{2cm}$ +\coeff{2}( q_{(u,v)}(t) 
            + q_{(v,u)}(t) )\big)\big\}$ \\
        \STATE $x'_{p^{*}}(t)\leftarrow x'_{p^{*}}(t)+1$
        \STATE $q_{(u,v)}(t) \leftarrow q_{(u,v)}(t)+\mathbbm{1}_{\{(u,v) \in p^{*}\}}-s_{(u,v)}(t) \ \forall (u,v) \in E$
     \STATE $a_{ij}(t) \leftarrow a_{ij}(t)-1$
        \ENDWHILE
        \IF{$\forall (u,v) \in E : q_{(u,v)}(t)\leq M_{(u,v)}$}
        \STATE $\mathbf{x}(t)=\mathbf{x}'(t)$, $\mathbf{x}'(t)=\mathbf{0}$
        \ELSE
        \STATE \COMMENT{On-chain Rebalancing}
        \STATE $r_{(i,j)}(t)=\begin{cases}
           a_{ij}(t) & \textit{w.p. } \delta \\
           0 & \textit{w.p. } 1-\delta \end{cases}$
        \ENDIF
    \ENDFOR
\STATE \textbf{Output:} $\mathbf{x}(t),\mathbf{r}(t)$
\end{algorithmic}
\end{algorithm}

The algorithm can be understood as follows: Consider the threshold $M$ to be different for each link in the network and set it to the capacity of each link, that is $M=c_{(u,v)}$ for $(u,v) \in E$. Also, rather than processing \$1 on-chain rebalancing, if the transaction is invoking on-chain rebalancing, then with probability $\delta$ process the whole transaction using on-chain rebalancing or reject the transaction otherwise. In addition, we route a transaction using multiple paths which balances the network. Now, we have an algorithm which processes payment requests using either only off chain rebalancing or only on chain rebalancing. 

{The algorithm carries out the following steps: In the first step, the money is not enqueued in the queues but a dummy variable $\mathbf{x}'$ is updated. Using this dummy variable, the queue lengths are updated and checked if any of them becomes greater than $M$. If this doesn't happen, then the routing variable $\mathbf{x}$ is set equal to $\mathbf{x}'$, otherwise, $\mathbf{x}$ remains $0$ and the transaction is completed via on-chain rebalancing with probability $\delta$ and rejected with probability $(1-\delta)$. Note that the routing algorithm only outputs $\mathbf{x}$ and $\mathbf{r}$. All the other auxiliary variables are inconsequential.}

Now with this modified algorithm, the metric for the performance will be the fraction of transactions that are processed using on chain rebalancing and the fraction of transactions that are rejected. This metric can be argued to be related to the throughput optimality as follows: the tail probability $\mathbb{P}[q_{(u,v)} \geq c_{(u,v)}]$ for all $(u,v) \in E$ can be bounded using the expected queue length given by \eqref{eq: expectedqueuelength} which will give us the probability with which a transaction will be either rejected or on chain rebalancing will be invoked. Now, we can tune the parameters $\delta$ and $\alpha$ to minimize the appropriate objective. We will verify this in the next section by considering a special case and show that it has good performance in simulations.  Also, this modified algorithm conforms with the PCN protocols and it can be argued as follows:
\begin{itemize}
    \item Under this algorithm, the atomicity of the payments will be preserved. This can be seen as follows:
    whenever a path is chosen to route the incoming demand, it is either routed and processed in the same time epoch as $M=c_{(u,v)}$ for all $(u,v) \in E$ or it is rejected with probability $(1-\delta)$ or it is processed using on-chain rebalancing with probability $\delta$. Thus, it preserves the atomicity of the payments.
    \item  Also, note that, it is no longer a single hop system now as either all the single hops are processed in one time epoch as the payment is routed only if the capacity is available or the whole payment is rejected or processed through on chain rebalancing.
    \item As $\delta$ approaches 0, the frequency of on chain rebalancing will be reduced. Thus, on-chain rebalancing won't be carried out in each time epoch. This is desirable as on-chain rebalancing is slower compared to off chain rebalancing.
\end{itemize}

\section{Simulations \label{sec:simulation}}

In queueing systems, closed loop policies that use the state information to make scheduling decisions are known to have better performance than open loop policies. Our proposed Algorithm 1 is a closed loop policy since it used the state information, as opposed to Spider \cite{spider}, which is an open loop policy.  
In this section, we will compare the performance of  Algorithm 1 with that of Fluid routing using simulations. Later, we discuss that Spider \cite{spider} implements fluid routing.

To study the efficiency of the Algorithm 1, we will drop all the transactions which invokes on chain rebalancing in our algorithm i.e., we will drop any transaction which if added to the payment graph, will lead to at least one edge having weight greater than or equal to $M$. This will give us the percentage of transactions successful using only off-chain rebalancing. We assume all the weights have equal capacity same as $M$ for simplicity. We use this as a metric (as opposed to calculating the rate of on-chain rebalancing) in order to be consistent with the literature   \cite{spider, silentwhispers, speedy}.

If we pick $M=c_{(u,v)}$ for every edge in the network and drop all the demand which requires on chain rebalancing, then at any instance, the queue length for each edge is less than its capacity. As the edge has enough capacity to process any buffered transaction at any point of time, this can be thought of as a system with no buffered transactions waiting in the queue to be processed. So although our algorithm is a queuing model, it can be implemented in such a way that there are no transactions waiting in the system. Note that, here we pick $M$ to be different for different edges which is a generalization of the algorithm we presented and the proof of throughput optimality can be easily extended for this system. The codes to run Algorithm \ref{alg2} and \ref{alg: water_filling} is open-source \cite{codeblockchain}.
\subsection{Setup}
\textbf{Simulator:} We use Python (networkx package) to create the graph and simulate the transactions. Our simulator finds the path in the graph according to the {Algorithm \ref{alg2}} and drops all the transactions which require on-chain rebalacing. Then we add the transaction to the outstanding queues of the edges. This can be thought of as each node has a capacity of $M$ units and if a transaction is processed, it will reduce the capacity of those edges by adding it to the outstanding queue lengths. Also with each transaction that is accepted, $\mathbf{q}$ is updated.    

\textbf{Schemes:} We run different variants of our algorithm by varying the parameters involved. We vary $M$ and report the percentage of successful transactions by units. We also report the average imbalance per edge to show the trade off between $M$ and queue length. 

\textbf{Metrics:} The metrics of our study are the \textit{average imbalance per edge} and \textit{success ratio}. Success ratio captures the fraction of transactions that were successful. 

\textbf{Data:} In the next sub-section, we will consider an illustrative example on a 10 node network. In addition, we carry out more extensive simulations on ripple network with 59,000 nodes in the appendix.

\subsection{Synthetic Data}
We consider five graphs, each with 10 nodes and 15-20 edges, that are generated randomly. We compare the average success ratio of Algorithm 1 and 2, and `fluid routing'.

\textbf{Generation of Transactions:} We start by generating a doubly stochastic matrix which is in the capacity region $\mathcal{C}$. Using the generated matrix as the demand rate matrix, we generate a total of 150,000 Poisson distributed transactions. From these generated transactions, we discarded all the transactions that are of 0 value. The generated transactions are then used in a random order to simulate the system. 

\textbf{Implementation of Fluid Routing}: {To implement the fluid routing, throughput is maximized subject to the capacity constraints (fluid LP in \cite{spider} is solved) which gives us optimal flow rates through all the paths for routing fluid demand $\boldsymbol{\lambda}$. In the long run, the frequency of usage of the paths is then maintained proportional to the optimal flow rates. In particular, the fraction of time each schedule is used is done in a randomized manner. 
The Spider algorithm in \cite{spider} is an approximate implementation of such a fluid routing algorithm. 
Since the fluid routing algorithm makes decisions in a randomized manner, we present the success ratio averaged over ten sample paths. }

{\textbf{Implementation of Algorithm \ref{alg2}:} For each transaction request from $i$ to $j$, all the paths in the network is enumerated and its weight given by Algorithm \ref{alg2} is calculated. Then, the path with minimum weight is chosen. Then, we route the transaction, only if the chosen path has enough capacity.}

\textbf{Implementation of Algorithm \ref{alg: water_filling} (Waterfilling):} Each transaction request is divided into multiple packets of value 1 and Algorithm \ref{alg2} is implemented for each packet to find multiple paths to route the incoming transaction. If there is enough capacity on the chosen paths, the complete transaction is processed or the complete transaction is declined. This algorithm tries to balance the channels by dividing a large transaction into multiple smaller ones.

\begin{figure}[t]
\centerline{\includegraphics[width=0.49\textwidth]{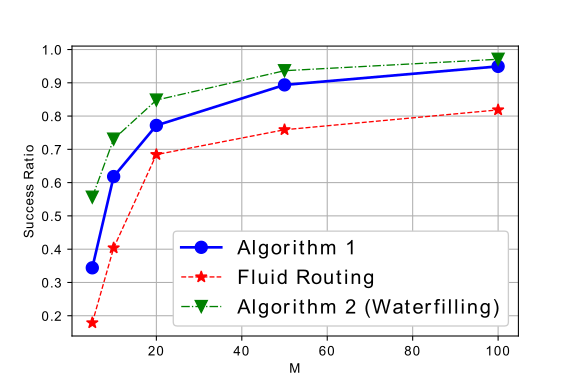}}
\caption{Comparison of Success ratio for Spider, Algorithm 1, and Algorithm \ref{alg: water_filling} for different $M$ on a 10 node graph}
\label{fig:mesh1}
\end{figure}

{Finally, also note that when comparing Algorithm \ref{alg2}, Algorithm \ref{alg: water_filling} and fluid routing, on chain rebalancing is not carried out in all the algorithms. In particular, whenever any of the algorithms invoke on chain rebalancing, the transaction is rejected and the effectiveness of the algorithms is compared using the success ration given in Fig. \ref{fig:mesh1}.}

Even if the exact transaction rate matrix is used (which cannot be found exactly in real life implementation), our algorithm consistently performs better than fluid routing for a wide range of values of $M$.  This is due to the fact that our algorithm actively tries to balance the graph whereas fluid routing aims to keep the graph balanced in the long run. This can be thought of as an open loop control whereas our algorithm makes decision based on the current state of the system and thus, is a closed loop control. In addition, Waterfilling improves the performance of Algorithm \ref{alg2} further as it routes an incoming transaction using multiple paths which ensures the graph is balanced.

We identify that the implementation of Algorithm \ref{alg2} and \ref{alg: water_filling} is NP-hard and cannot be implemented for large instances of P2P networks. We design heuristic implementation of these algorithms and compare it with exact implementation for the synthetic data. In addition, we also implement the heuristic on the real life data (ripple network) with 59,000 nodes and 191,242 edges. We show that the heuristic implementation achieves similar success ratio as the exact implementation for the synthetic data and the heuristic implementation performs well for the real life data. 

\subsection{Heuristic Implementation}
Algorithm \ref{alg2} and \ref{alg: water_filling} intends to find path(s) from a sender to receiver which is the most imbalanced and least loaded path. To implement this algorithm, we need to find the shortest path in a weighted directed graph with negative weights allowed. As this problem is known to be an NP-hard problem (it can be decomposed from Hamiltonian path problem \cite{garey2002computers}), we use a heuristic implementation of the algorithm for computational viability.

\begin{figure}[b]
  \centerline{\includegraphics[width=0.48\textwidth]{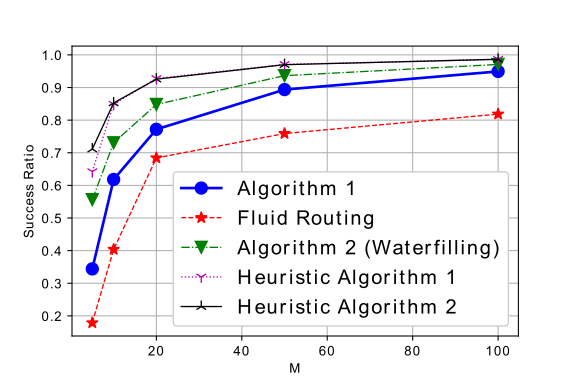}}
\caption{Comparison of Success ratio for Spider, Algorithm 1, and Algorithm \ref{alg: water_filling} for different $M$ on a 10 node graph}
\label{fig: heu}
\end{figure}
\begin{figure}[t]
  \centerline{\includegraphics[width=0.48\textwidth]{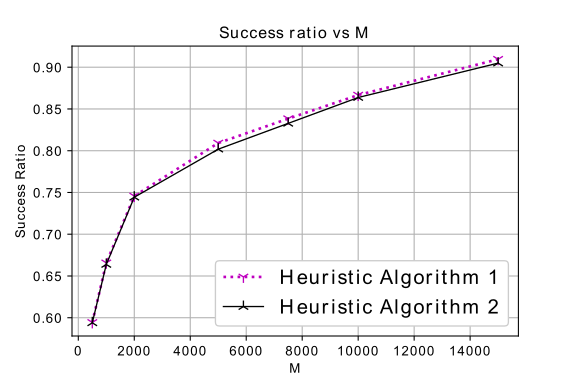}}
\caption{Success ratio of algorithm 1 for the ripple network plotted against the maximum allowable imbalance (capacity: $M$)}
\label{fig: sucess-ratio}
\end{figure}

\textbf{Implementation of Heuristic Version of Algorithm \ref{alg2}:}
The heuristic we implement for the real life data is to find k-shortest paths \cite{kshortestpath} from the sender to receiver by considering only the positive part of the weight $[z_{(u,v)}+\coeff{2}(q_{(u,v)}+q_{(v,u)})]^{+}$ on each edge. Thus, it becomes a problem of finding the shortest path in a weighted directed graph with all the weights non-negative. From those k- shortest path, we will choose the path with minimum actual weight, i.e. the path such that the sum of the weights $z_{(u,v)}+\coeff{2}(q_{(u,v)}+q_{(v,u)})$, for all $(u,v) \in p_{ij}$ is minimized. Here, $i$ is the source and $j$ is the destination of the arising transaction. Algorithm \ref{alg2} routes the incoming transaction on this path if it has enough capacity.

If k is equal to the total possible paths between a sender and receiver, then the heuristic implementation becomes the exact implementation of the algorithm. The complexity of the well known algorithm \cite{kshortestpath} to find $k$ shortest loopless paths in a network is pseduo-polynomial given by $O\left(k|V|\left(|E|+|V|\log|V|\right)\right)$. In addition to this, we require a maximum of $O(|E|)$ comparisons to make the on-chain rebalancing decisions.

\textbf{Implementation of Heuristic Version of Algorithm \ref{alg: water_filling}:} Each incoming transaction is divided into certain number of packets and a path for each packet is computed by implementing the heuristic implementation of Algorithm \ref{alg2}. If there is enough capacity on the chosen paths, the complete transaction is processed or the complete transaction is declined.

Finally, note that we pick $k=1$ for the simulations, i.e. we route the incoming transaction on the shortest weighted path with  $[z_{(u,v)}+\coeff{2}(q_{(u,v)}+q_{(v,u)})]^{+}$ as the weight for the edge $(u,v) \in E$. In addition, we divide each incoming transaction into 2 packets to implement the Algorithm \ref{alg: water_filling}.

As observed in Fig. \ref{fig: heu}, the heuristic implementation of both the algorithms matches closely with the exact versions for large values of $M$. Rather surprisingly,  the heuristic implementation leads to a better success ratio for small values of $M$. A possible explanation is that since the heuristic simply drops negative weights, it chooses  shorter paths compared to an exact implementation. Investigating the throughput optimality of the heuristic algorithms is a direction of  future work. To motivate the heuristics further, in the next section, we will implement them on a real life data set.
\begin{figure}[b]
    \centering
\centerline{\includegraphics[width=0.48\textwidth]{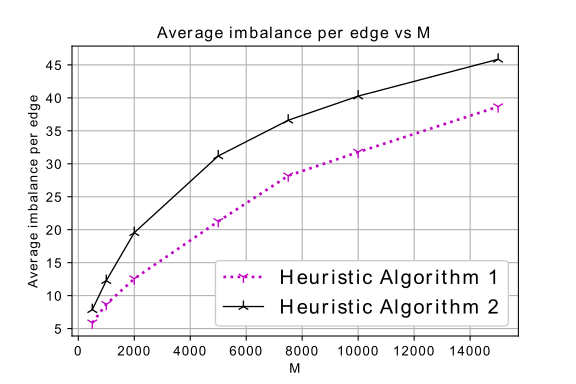}}
\caption{Imbalance per edge in the graph after 50,000 transactions routed through the network using algorithm \ref{alg2}}
\label{fig: imbalance}
\end{figure}
\begin{figure}[t]
    \centerline{\includegraphics[width=0.48\textwidth]{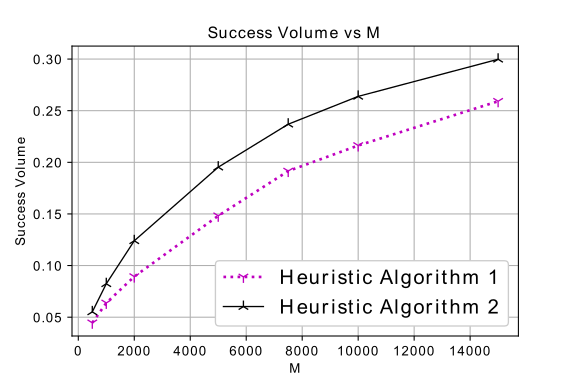}}
\caption{Success volume of algorithm 1 for the ripple network plotted against the maximum allowable imbalance (capacity: $M$)}
\label{fig: success-volume}
\end{figure}
\subsection{Simulations: Ripple Network} \label{app: simulations}
We run the heuristic implementation of Algorithm \ref{alg2} and heuristic implementation of water-filling algorithm \ref{alg: water_filling} as explained in the previous sub-section and compare the results obtained in the two cases. In this section, we pick $k=20$ and divide each incoming transaction into 5 packets for Algorithm \ref{alg: water_filling}.

\textbf{Data set:} In this section, we use the data-set provided by Speedy Murmurs \cite{speedysoftware} which is a sub-graph from the original topology of Ripple. The data-set consists of 59,000 nodes and 191,242 edges. Speedy Murmurs removed the inconsistencies from the original Ripple Data and it was filtered and transformed into a format which was convenient to run simulations on. We are grateful to the authors for providing the data-set. {We simulated the system using 41,491 transactions with an average transaction amount of 415 units, median of 1.314 units and a maximum transaction of $500$ units. These are real transactions occurred in the ripple network and provided by Speedy Murmurs. In particular, we took the data set of 50,000 transactions provided by SpeedyMurmurs and discarded all the transactions that were above 500 units as these transactions cannot be fulfilled if $M=500$.}


\textbf{Results}
    With the increasing capacity of the edges, the throughput in terms of success ratio increases. The increasing success ratio with the capacity $M$ is very intuitive as with the increase in capacity of the edges, the system state is allowed to buffer more transactions per edge and thus, more transactions will be processed. {For large enough $M$, Algorithm \ref{alg2} and \ref{alg: water_filling} are able to route most of the transaction requests successfully.}

The average imbalance per edge increases with the increase in the capacity as with the increasing capacity, we are allowing more buffered transactions which results in the increase of the average imbalance per edge in the steady state. {Although, the average imbalance stays quite low which implies that our algorithm does a good job of routing using the paths which leads to balanced edges. By theorem \ref{theorem}, even though, we route all the transactions and buffer them until they are met, the average imbalance per edge will remain finite as the underlying Markov chain is positive recurrent.}

With the increasing success ratio, the success volume also increases which can be seen in Fig.~\ref{fig: success-volume}. {Water-filling performs much better than Algorithm \ref{alg2} in terms of success volume as it is using multiple routes, thus, higher valued transactions are split into multiple transactions and processed successfully.}

The results are summarized in Fig.~\ref{fig: sucess-ratio}, Fig.~\ref{fig: imbalance} and Fig.~\ref{fig: success-volume}. 

It is evident by the results that Water-Filling algorithm performs better than Algorithm \ref{alg2} as it divided the incoming transaction into multiple routes to balance the imbalance across edges.

\section{Conclusion and Future Work \label{sec:conclusion}}
In this paper, we presented a novel stochastic model to study the routing of payment requests in a Blockchain based payment processing networks. 
We defined two notions of stability and then characterized the capacity region of the system consisting of all the demand rates under which the system is stable. We then presented a novel MaxWeight like state dependent algorithm and proved that it is throughput optimal. We showed that the rate of on chain rebalancing used is $O(\delta)$ and the expected sum of queue lengths under the stationary distribution of the DTMC is $O(1/\delta^2)$. 
We argued that a state dependent policy has a better performance than open loop policy by simulating our algorithm and compare it with Spider \cite{spider}, which uses an open loop policy. 

Since the proposed algorithm is NP-hard in general, finding a low-complexity but closed loop algorithm that is throughput optimal is future work. Another future direction is to study throughput maximizing routing algorithms that can be implemented in a distributed manner so that privacy of the transactions is preserved.

\bibliographystyle{IEEEtran}
\bibliography{references}

\begin{thebibliography}{10}
\providecommand{\url}[1]{#1}
\csname url@samestyle\endcsname
\providecommand{\newblock}{\relax}
\providecommand{\bibinfo}[2]{#2}
\providecommand{\BIBentrySTDinterwordspacing}{\spaceskip=0pt\relax}
\providecommand{\BIBentryALTinterwordstretchfactor}{4}
\providecommand{\BIBentryALTinterwordspacing}{\spaceskip=\fontdimen2\font plus
\BIBentryALTinterwordstretchfactor\fontdimen3\font minus
  \fontdimen4\font\relax}
\providecommand{\BIBforeignlanguage}[2]{{%
\expandafter\ifx\csname l@#1\endcsname\relax
\typeout{** WARNING: IEEEtran.bst: No hyphenation pattern has been}%
\typeout{** loaded for the language `#1'. Using the pattern for}%
\typeout{** the default language instead.}%
\else
\language=\csname l@#1\endcsname
\fi
#2}}
\providecommand{\BIBdecl}{\relax}
\BIBdecl

\bibitem{lightningnetwork}
J.~Poon and T.~Dryja, ``The bitcoin lightning network: Scalable off-chain
  instant payments,'' 2016.

\bibitem{spider}
V.~Sivaraman, S.~B. Venkatakrishnan, K.~Ruan, P.~Negi, L.~Yang, R.~Mittal,
  M.~Alizadeh, and G.~Fanti, ``High throughput cryptocurrency routing in
  payment channel networks,'' 2018.

\bibitem{satoshibitcoin}
S.~Nakamoto, ``Bitcoin: A peer-to-peer electronic cash system,'' 2008.

\bibitem{wood2014ethereum}
G.~Wood, ``Ethereum: A secure decentralised generalised transaction ledger,''
  \emph{Ethereum project yellow paper}, vol. 151, pp. 1--32, 2014.

\bibitem{smartcontract}
V.~Buterin \emph{et~al.}, ``A next-generation smart contract and decentralized
  application platform,'' \emph{white paper}, vol.~3, p.~37, 2014.

\bibitem{smartcontract2}
P.~Franco, \emph{Understanding bitcoin}.\hskip 1em plus 0.5em minus 0.4em\relax
  Wiley Online Library, 2014.

\bibitem{blockchainbanking}
Y.~Guo and C.~Liang, ``Blockchain application and outlook in the banking
  industry,'' \emph{Financial Innovation}, vol.~2, no.~1, p.~24, 2016.

\bibitem{nash2016ibm}
K.~S. Nash, ``{IBM} pushes blockchain into the supply chain,'' The Wall Street
  Journal. Available online: https://www. wsj.
  com/articles/ibm-pushes-blockchain-into-the-supplychain-1468528824, 2016.

\bibitem{marvin2017blockchain}
R.~Marvin, ``Blockchain: The invisible technology that's changing the world.
  {PCM}ag,'' 2017.

\bibitem{hybridblockchain}
L.~Wu, K.~Meng, S.~Xu, S.~Li, M.~Ding, and Y.~Suo, ``Democratic centralism: A
  hybrid blockchain architecture and its applications in energy internet,'' in
  \emph{2017 IEEE International Conference on Energy Internet (ICEI)}.\hskip
  1em plus 0.5em minus 0.4em\relax
  https://ieeexplore.ieee.org/document/7926870: IEEE, 2017, pp. 176--181.

\bibitem{ghosh2007mechanism}
A.~Ghosh, M.~Mahdian, D.~M. Reeves, D.~M. Pennock, and R.~Fugger, ``Mechanism
  design on trust networks,'' in \emph{International Workshop on Web and
  Internet Economics}.\hskip 1em plus 0.5em minus 0.4em\relax
  https://dl.acm.org/citation.cfm?id=1781925: Springer, 2007, pp. 257--268.

\bibitem{rippleoverview}
F.~Armknecht, G.~O. Karame, A.~Mandal, F.~Youssef, and E.~Zenner, ``Ripple:
  Overview and outlook,'' in \emph{International Conference on Trust and
  Trustworthy Computing}.\hskip 1em plus 0.5em minus 0.4em\relax Springer,
  2015, pp. 163--180.

\bibitem{bitcoinvsvisa}
J.~Vermeulen, ``Bitcoin and ethereum vs visa and paypal--transactions per
  second,'' \emph{My Broadband}, vol.~22, 2017.

\bibitem{ppcoinstake}
S.~King and S.~Nadal, ``Ppcoin: Peer-to-peer crypto-currency with
  proof-of-stake,'' \emph{self-published paper, August}, vol.~19, 2012.

\bibitem{scalingblockchain}
K.~Croman, C.~Decker, I.~Eyal, A.~E. Gencer, A.~Juels, A.~Kosba, A.~Miller,
  P.~Saxena, E.~Shi, E.~G. Sirer \emph{et~al.}, ``On scaling decentralized
  blockchains,'' in \emph{International Conference on Financial Cryptography
  and Data Security}.\hskip 1em plus 0.5em minus 0.4em\relax Springer, 2016,
  pp. 106--125.

\bibitem{raiden}
R.~Mitra, ``Lightning protocol \& the raiden network: a beginner’s guide,''
  \emph{Retrieved October}, vol.~26, p. 2018, 2017.

\bibitem{atomicswap}
M.~Herlihy, ``Atomic cross-chain swaps,'' in \emph{Proceedings of the 2018 ACM
  Symposium on Principles of Distributed Computing}.\hskip 1em plus 0.5em minus
  0.4em\relax https://dl.acm.org/citation.cfm?id=3212736: ACM, 2018, pp.
  245--254.

\bibitem{silentwhispers}
P.~Moreno-Sanchez, A.~Kate, and M.~Maffei, ``Silentwhispers: Enforcing security
  and privacy in decentralized credit networks,'' in \emph{24th Network and
  Distributed System Security Symposium (NDSS 2018)}, 2017.

\bibitem{speedy}
S.~Roos, P.~Moreno-Sanchez, A.~Kate, and I.~Goldberg, ``Settling payments fast
  and private: Efficient decentralized routing for path-based transactions,''
  in \emph{Proceedings 2018 Network and Distributed System Security
  Symposium}.\hskip 1em plus 0.5em minus 0.4em\relax
  http://dx.doi.org/10.14722/ndss.2018.23252: NDSS, 01 2018.

\bibitem{ganjali2005cell}
Y.~Ganjali, A.~Keshavarzian, and D.~Shah, ``Cell switching versus packet
  switching in input-queued switches,'' \emph{IEEE/ACM Transactions on
  Networking (TON)}, vol.~13, no.~4, pp. 782--789, 2005.

\bibitem{wu2016privacy}
D.~J. Wu, J.~Zimmerman, J.~Planul, and J.~C. Mitchell, ``Privacy-preserving
  shortest path computation,'' \emph{arXiv preprint arXiv:1601.02281}, 2016.

\bibitem{tsuchiya1988landmark}
P.~F. Tsuchiya, ``The landmark hierarchy: a new hierarchy for routing in very
  large networks,'' in \emph{ACM SIGCOMM Computer Communication Review},
  vol.~18.\hskip 1em plus 0.5em minus 0.4em\relax
  https://dl.acm.org/citation.cfm?id=52329: ACM, 1988, pp. 35--42.

\bibitem{matchingqueues}
I.~Gurvich and A.~Ward, ``On the dynamic control of matching queues,''
  \emph{Stochastic Systems}, vol.~4, no.~2, pp. 479--523, 2014.

\bibitem{banerjeeridehailing}
\BIBentryALTinterwordspacing
S.~Banerjee, D.~Freund, and T.~Lykouris, ``Pricing and optimization in shared
  vehicle systems: An approximation framework,'' in \emph{Proceedings of the
  2017 ACM Conference on Economics and Computation}, ser. EC '17.\hskip 1em
  plus 0.5em minus 0.4em\relax New York, NY, USA: ACM, 2017, pp. 517--517.
  [Online]. Available: \url{http://doi.acm.org/10.1145/3033274.3085099}
\BIBentrySTDinterwordspacing

\bibitem{yashkanoriaridehailing}
S.~Banerjee, Y.~Kanoria, and P.~Qian, ``State dependent control of closed
  queueing networks with application to ride-hailing,'' arXiv preprint
  arXiv:1803.04959, 2018.

\bibitem{ondemandservers}
L.~M. Nguyen and A.~L. Stolyar, ``A queueing system with on-demand servers:
  local stability of fluid limits,'' \emph{Queueing Systems}, vol.~89, no. 3-4,
  pp. 243--268, 2018.

\bibitem{caldentey2009fcfs}
R.~Caldentey, E.~H. Kaplan, and G.~Weiss, ``{FCFS} infinite bipartite matching
  of servers and customers,'' \emph{Advances in Applied Probability}, vol.~41,
  no.~3, pp. 695--730, 2009.

\bibitem{adan2012exact}
I.~Adan and G.~Weiss, ``Exact {FCFS} matching rates for two infinite multitype
  sequences,'' \emph{Operations research}, vol.~60, no.~2, pp. 475--489, 2012.

\bibitem{roth2007kidney}
A.~E. Roth, T.~S{\"o}nmez, and M.~U. {\"U}nver, ``Efficient kidney exchange:
  Coincidence of wants in markets with compatibility-based preferences,''
  \emph{American Economic Review}, vol.~97, no.~3, pp. 828--851, 2007.

\bibitem{yashkanoriabarter}
R.~Anderson, I.~Ashlagi, D.~Gamarnik, and Y.~Kanoria, ``Efficient dynamic
  barter exchange,'' \emph{Operations Research}, vol.~65, no.~6, pp.
  1446--1459, 2017.

\bibitem{akbarpour2017thickness}
M.~Akbarpour, S.~Li, and S.~Oveis~Gharan, ``Thickness and information in
  dynamic matching markets,'' \emph{Journal of Political Economy}, vol.
  forthcoming, 2019.

\bibitem{tassiulas1990stability}
L.~Tassiulas and A.~Ephremides, ``Stability properties of constrained queueing
  systems and scheduling policies for maximum throughput in multihop radio
  networks,'' in \emph{29th IEEE Conference on Decision and Control}.\hskip 1em
  plus 0.5em minus 0.4em\relax https://ieeexplore.ieee.org/document/182479:
  IEEE, 1990, pp. 2130--2132.

\bibitem{srikantbook}
R.~Srikant and L.~Ying, \emph{Communication networks: an optimization, control,
  and stochastic networks perspective}.\hskip 1em plus 0.5em minus 0.4em\relax
  Cambridge University Press, 2013.

\bibitem{maguluri2016heavy}
S.~T. Maguluri and R.~Srikant, ``Heavy traffic queue length behavior in a
  switch under the {M}ax{W}eight algorithm,'' \emph{Stochastic Systems},
  vol.~6, no.~1, pp. 211--250, 2016.

\bibitem{srikantad}
B.~Tan and R.~Srikant, ``Online advertisement, optimization and stochastic
  networks,'' \emph{IEEE Transactions on Automatic Control}, vol.~57, no.~11,
  pp. 2854--2868, 2012.

\bibitem{kanoria2019backpressure}
Y.~Kanoria and P.~Qian, ``Near optimal control of a ride-hailing platform via
  mirror backpressure,'' 2019.

\bibitem{satoshiunit}
R.~L. Twesige, ``A simple explanation of bitcoin and blockchain technology,''
  2015.

\bibitem{ratestability}
Y.~Ganjali, A.~Keshavarzian, and D.~Shah, ``Cell switching versus packet
  switching in input-queued switches,'' \emph{IEEE/ACM Transactions on
  Networking (TON)}, vol.~13, no.~4, pp. 782--789, 2005.

\bibitem{HajekComm}
B.~Hajek, ``Notes for ece 467 communication network analysis,'' December 2006.

\bibitem{codeblockchain}
S.~M. Varma and S.~T. Maguluri, ``Throughput optimal routing in blockchain
  based payment systems (codes)
  \url{https://github.com/gt-coar/Blockchain_P2P},'' 2020.

\bibitem{garey2002computers}
M.~R. Garey and D.~S. Johnson, \emph{Computers and intractability (page
  213)}.\hskip 1em plus 0.5em minus 0.4em\relax wh freeman New York, 2002,
  vol.~29.

\bibitem{kshortestpath}
J.~Y. Yen, ``Finding the k shortest loopless paths in a network,''
  \emph{management Science}, vol.~17, no.~11, pp. 712--716, 1971.

\bibitem{speedysoftware}
S.~Roos, P.~Moreno-Sanchez, A.~Kate, and I.~Goldberg, ``Speedymurmurs: Fast and
  private path-based transactions (software)
  https://crysp.uwaterloo.ca/software/speedymurmurs/,'' 2018.

\end{thebibliography}


\appendix
\subsection{Proof of Proposition \ref{prop}}\label{app:proofofprop}
\begin{proof}
As $\boldsymbol{\lambda} \notin \mathcal{C}$, there does not exist $\mathbf{x} \in \mathbb{R}_{+}^{|\mathcal{P}|}$ which satisfies the set of inequalities \eqref{prop1}, \eqref{prop2} and \eqref{prop3}. The dual of the linear program with objective function $\equiv 0$ and the constraints \eqref{prop1}, \eqref{prop2} and \eqref{prop3} will be  \eqref{eq: obj-dual}, \eqref{constrant1} and \eqref{constraint2}. As the primal is infeasible and dual is feasible (trivial feasible solution is $\boldsymbol{\zeta}=\mathbf{0}$,$\boldsymbol{\beta}=\mathbf{0}$,$\boldsymbol{\gamma}=\mathbf{0}$), the optimal dual value should be $-\infty$. 
Hence, there exists a  $\boldsymbol{\beta}^{*}$, $\boldsymbol{\gamma}^{*}$ and
\begin{align}
    \zeta_{ij}^{*}=\hspace{-0.1cm}-\hspace{-0.1cm}\min_{p \in P_{ij}}\hspace{-0.2cm} \sum_{(u,v) \in p}(\beta_{(u,v)}^{*}&+\beta_{(v,u)}^{*}+\gamma_{(u,v)}^{*}-\gamma_{(v,u)}^{*}) \forall i,j \in V, \nonumber
\end{align}
such that the objective function value of the dual $\eqref{eq: obj-dual}$ is equal to $-D$ (as we pick $\boldsymbol{\zeta}$ to be the minimum possible value in terms of $\boldsymbol{\beta}$ and $\boldsymbol{\gamma}$) for some $D>0$ i.e.,
\begin{align}
    -\sum_{i \neq j \in V}& \lambda_{ij}\min_{p \in P_{ij}}\hspace{-0.2cm} \sum_{(u,v) \in p}(\beta_{(u,v)}^{*}+\beta_{(v,u)}^{*}+\gamma_{(u,v)}^{*}-\gamma_{(v,u)}^{*}) \nonumber \\
    &+2\sum_{(u,v) \in E} c_{(u,v)}\beta_{(u,v)}^{*}=-D. \label{eq: prop1dualisunbounded}
\end{align}
Note that it is not possible to have $(\beta^{*}_{(u,v)}+\beta^{*}_{(v,u)}+\gamma^{*}_{(u,v)}-\gamma^{*}_{(v,u)})=0$ for all $(u,v) \in E$ and satisfy the above inequality. Thus, $\boldsymbol{\beta}^{*}$ and $\boldsymbol{\gamma}^{*}$ are such that $\max_{(u,v) \in E} \{\beta^{*}_{(u,v)}+\beta^{*}_{(v,u)}+\gamma^{*}_{(u,v)}-\gamma^{*}_{(v,u)}\} \neq 0$. Now, consider any family of routing algorithms parametrized by $\delta$ for the payment processing network. We will now calculate the weighted sum of all the queues and show that it is not finite. Consider the following:
\begin{align}
    &\sum_{(u,v) \in E} (\beta_{(u,v)}^{*}+\beta_{(v,u)}^{*}+\gamma_{(u,v)}^{*}-\gamma_{(v,u)}^{*})q_{(u,v)}(T)  \label{eq: prop1weightedsumofqueuelengths} \\
     \overset{*}{=}{}&\sum_{(u,v) \in E}\hspace{-2.5pt}\bigg[ (\beta_{(u,v)}^{*}+\beta_{(v,u)}^{*}+\gamma_{(u,v)}^{*}-\gamma_{(v,u)}^{*})q_{(u,v)}(0)\nonumber \\ &+\sum_{t=1}^T(\beta_{(u,v)}^{*}+\beta_{(v,u)}^{*}+\gamma_{(u,v)}^{*}-\gamma_{(v,u)}^{*})\big(y_{(u,v)}(t)\nonumber\\
     &-s_{(u,v)}(t)-r_{(u,v)}(t)\big)\bigg] \nonumber \\
    \overset{**}{\geq}{}&  \sum_{(u,v) \in E}\hspace{-2.5pt}\bigg[ (\beta_{(u,v)}^{*}+\beta_{(v,u)}^{*}+\gamma_{(u,v)}^{*}-\gamma_{(v,u)}^{*})q_{(u,v)}(0)\nonumber\\
    &+\sum_{t=1}^T(\beta_{(u,v)}^{*}+\beta_{(v,u)}^{*}+\gamma_{(u,v)}^{*}-\gamma_{(v,u)}^{*})\big(y_{(u,v)}(t)\nonumber\\
    &-r_{(u,v)}(t)\big)-2\beta_{(u,v)}^{*}c_{(u,v)}(t)\bigg], \label{eq: prop1-sumofquv}
\end{align}
where $(*)$ follows from the telescopic sum of \eqref{qu} and, $(**)$ follows from the definition of $s_{(u,v)}$ given in \eqref{eq:suv} and as $s_{(u,v)}=s_{(v,u)}$, we have $\sum_{(u,v) \in E}s_{(u,v)}(t)\gamma_{(u,v)}^{*}=\sum_{(u,v) \in E}s_{(u,v)}(t)\gamma_{(v,u)}^{*}$. As $\boldsymbol{\beta}^{*}$ is non negative, we can then replace $s_{(u,v)}(t)$ by its upper bound $c_{(u,v)}$ and use $c_{(u,v)}=c_{(v,u)}$ to write \eqref{eq: prop1-sumofquv}. Now we will substitute $\mathbf{y}$ in terms of $\mathbf{a}$. Consider the following: $\sum_{(u,v)}w_{(u,v)}(t)y_{(u,v)}(t)$ is the weighted sum of total demand that is routed, where the demand routed through the edge $(u,v)$ is given a weight $w_{(u,v)}$. Now, this is exactly equal to the total demand that is routed ($a_{ij}(t)$ for all $i,j \in V$) using a certain set of paths $\mathbf{p}(t)$, where the weight given to every path is the sum of the weights of the edges in the path, i.e. $\sum_{(u,v) \in p} w_{(u,v)}$. So we have,
\begin{align}
    \lefteqn{\sum_{(u,v) \in E} \hspace{-5pt} w_{(u,v)}y_{(u,v)}(t)} \nonumber\\
    &=\sum_{(u,v) \in E} \hspace{-5pt} w_{(u,v)}\sum_{i\neq j \in V} \sum_{p \in P_{ij}} x_{p}(t)\mathbbm{1}_{(u,v) \in p}\nonumber \\
    &=\sum_{i\neq j \in V}\sum_{p \in P_{ij}} x_p(t)\sum_{(u,v) \in p} w_{(u,v)} \nonumber \\
    &=\sum_{i\neq j \in V} a_{ij}(t)\sum_{p \in P_{ij}}\alpha_{p}^{ij}(t)\sum_{(u,v) \in p} w_{(u,v)} \nonumber \\
    & \overset{*}{\geq} \hspace{-3pt}\sum_{(i,j \in V)}a_{ij}(t)\min_{p \in P_{ij}}\sum_{(u,v) \in p}w_{(u,v)}, \label{eq: proof-prop1-keystep}
\end{align}
where $\sum_{p \in P_{ij}}\alpha_p^{ij}(t)=1$ and $\alpha_p^{ij}(t) \geq 0$ for all $i\neq j \in V$ and it is the fraction of incoming demand from $i$ to $j$ routed through path $p$, i.e. $x_p(t)=\alpha_{p}^{ij}(t)a_{ij}(t)$. Also, $(*)$ follows as minimum of a set is always less than any convex combination of the elements of a set.

Note that, this is the key step in the proof of this proposition as we now have substituted $\mathbf{y}$ in terms of $\mathbf{a}$ which is assumed to be i.id across time and vertices.

Now we will show that \eqref{eq: prop1-sumofquv} divided by $T$ is greater than zero as follows:
\begin{align}
    &\frac{1}{T}\sum_{(u,v) \in E}(\beta_{(u,v)}^{*}+\beta_{(v,u)}^{*}+\gamma_{(u,v)}^{*}-\gamma_{(v,u)}^{*})q_{(u,v)}(0)\nonumber \\
    &+\frac{1}{T}\sum_{t=1}^T\sum_{(u,v) \in E}\bigg[(\beta_{(u,v)}^{*}+\beta_{(v,u)}^{*}+ \gamma_{(u,v)}^{*}-\gamma_{(v,u)}^{*}) \times \nonumber \\ &\big(y_{(u,v)}(t)-r_{(u,v)}(t)\big)-2\beta_{(u,v)}^{*}c_{(u,v)}(t)\bigg] \nonumber \\
    \overset{*}{\geq}{}& \frac{1}{T}\hspace{-7pt}\sum_{(u,v) \in E}\hspace{-3pt}(\beta_{(u,v)}^{*}+\beta_{(v,u)}^{*}+\gamma_{(u,v)}^{*}-\gamma_{(v,u)}^{*})q_{(u,v)}(0)\nonumber\\
    &+\frac{1}{T}\sum_{t=1}^T \bigg[ \sum_{i \neq j \in V} a_{ij}(t)\min_{p \in P_{ij}}\hspace{-3pt}\bigg\{\hspace{-3pt}\sum_{(u,v) \in E}(\beta_{(u,v)}^{*}+\beta_{(v,u)}^{*}\nonumber \\
    &+\gamma_{(u,v)}^{*}-\gamma_{(v,u)}^{*})\bigg\}-\sum_{(u,v) \in E}(\beta_{(u,v)}^{*}+\beta_{(v,u)}^{*}+\gamma_{(u,v)}^{*}  \nonumber \\ &-\gamma_{(v,u)}^{*})r_{(u,v)}(t)\bigg]-2\sum_{(u,v)\in E}\beta_{(u,v)}^{*}c_{(u,v)}(t), \label{eq: prop1yuvtoaij}
\end{align}
where $(*)$ follows from \eqref{eq: proof-prop1-keystep}.  Thus, we can now take the $\limsup$ as $T \rightarrow \infty$ of \eqref{eq: prop1yuvtoaij} and use strong law of large numbers for i.id random variables to get: 
\begin{align}
    \frac{\eqref{eq: prop1weightedsumofqueuelengths}}{T}\overset{*}{\geq}{}& \sum_{i \neq j \in V} \lambda_{ij}\min_{p \in P_{ij}}\bigg\{ \hspace{-2pt}\sum_{(u,v) \in p}(\beta_{(u,v)}^{*}+\beta_{(v,u)}^{*}+\gamma_{(u,v)}^{*} \nonumber \\
    & -\gamma_{(v,u)}^{*})\bigg\}-2\sum_{(u,v) \in E}\beta_{(u,v)}^{*}c_{(u,v)}-B^{'}\delta \textit{ w.p. 1} \nonumber \\
    \overset{**}{=}{}& (D-B^{'}\delta) >0  \textit{ w.p. 1}\ \forall \delta < \frac{D}{B^{'}} \label{eq: prop1geq0}
\end{align}
where $B^{'}=\max_{(u,v) \in E}[\beta_{(u,v)}^{*}+\beta_{(v,u)}^{*}+\gamma_{(u,v)}^{*}-\gamma_{(v,u)}^{*}]^{+}$ as $r_{(u,v)}(t) \geq 0$ for any $t$, $(u,v) \in E$. Also, as $\limsup_{T \rightarrow \infty} \frac{1}{T}\sum_{t=1}^T\sum_{(u,v) \in E} r_{(u,v)}(t) \leq \delta$, $(*)$ follows. Further, $(**)$ follows from \eqref{eq: prop1dualisunbounded}.

By using \eqref{eq: prop1geq0}, for on chain rebalancing rate ($\delta$) less than $\frac{D}{B^{'}}$, we have $\sum_{(u,v) \in E} (\beta_{(u,v)}^{*}+\beta_{(v,u)}^{*}+\gamma_{(u,v)}^{*}-\gamma_{(v,u)}^{*})q_{(u,v)}(T) \rightarrow \infty$ with probability 1. As the weighted sum (with at least one weight non zero) of the queue lengths tends to infinity with probability 1, we have:
\begin{align}
    \lim_{D \rightarrow \infty} \lim_{t \rightarrow \infty}\mathbb{P}\left(\sum_{(u,v) \in E} q_{(u,v)}(t) \geq D\right)=1 \ \forall \delta < \frac{D}{B ^{'}}.
\end{align}
Thus, no family of algorithm can weakly stabilize the system if the demand matrix is outside the capacity region $\mathcal{C}$.
\end{proof}
\section{Proof of Theorem \ref{theorem}} \label{appendix: theorem}
\subsection{Proof of Theorem \ref{theorem}}
\begin{proof}[\textbf{Proof of Theorem 1}]
Defining the effective demand in one time epoch for each edge for the ease of the presentation as,
\begin{equation} \label{y}
    y_{(u,v)}=\sum_{i\neq j \in V}\sum_{p \in P_{ij}} x_p \mathbbm{1}_{(u,v) \in p}.
\end{equation}
Note that, this is equivalent to the definition of $y_{(u,v)}$ as defined in \eqref{eq: yuv}. We will now define the Lyapunov function $V(\xs)=V_1(\xs)+V_2(\xs)$ to be a quadratic function of the queue lengths as,
\begin{align} 
    V_1(\xs)&\overset{\Delta}{=}\sum_{(u,v) \in E} z_{(u,v)}^2, \label{lyapunov1} \\
    V_2(\xs)&\overset{\Delta}{=}\sum_{(u,v) \in E} \coeff{2}(q_{(u,v)}+q_{(v,u)})^2. \label{lyapunov2}
\end{align}
The queue length $q_{(u,v)}$ evolves according to \eqref{qu} and the imbalance $z_{(u,v)}$ evolves as,
\begin{align}
     z_{(u,v)}(t+1)={}&z_{(u,v)}(t)+y_{(u,v)}(t)-y_{(v,u)}(t) \nonumber \\
     &-r_{(u,v)}(t)+r_{(v,u)}(t). \label{z}
\end{align}
For the simplicity of notation, we will denote $\mathbb{E}[.|\mathbf{q}(t)=\mathbf{q}]$ by $\mathbb{E}_{\xs}[.]$. Now we will separately calculate the drift of $V_1$ and $V_2$ and then add them to calculate the drift of $V$ as,
\begin{align*}
    \lefteqn{\mathbb{E}_{\xs}[\Delta V_1]} \\ 
    ={}&\mathbb{E}_{\xs}\bigg[\sum_{(u,v) \in E}\left(z_{(u,v)}^2(t+1)-z_{(u,v)}^2(t)\right)\bigg]  \\
    ={}& \mathbb{E}_{\xs}\bigg[\sum_{(u,v) \in E}\hspace{-0.85pt}\left(y_{(u,v)}(t)-y_{(v,u)}(t)-r_{(u,v)}(t)+r_{(v,u)}(t)\right)^2  \nonumber \\
    & \hspace{-11pt}+\hspace{-2pt}2\hspace{-8pt}\sum_{(u,v) \in E} \hspace{-8pt} z_{(u,v)}(t)(y_{(u,v)}(t)-y_{(v,u)}(t)-r_{(u,v)}(t)+r_{(v,u)}(t))\bigg] \numberthis \label{vsim1}
\end{align*}
We will now show that the first term in \eqref{vsim1} is bounded. We have:
\begin{align}
    T_1(\xs)&=\mathbb{E}_{\xs}\bigg[\hspace{-3.5pt}\sum_{(u,v) \in E}\hspace{-9pt}\left(y_{(u,v)}(t)\hspace{-2.25pt}-\hspace{-2.25pt}y_{(v,u)}(t)\hspace{-2.25pt}-\hspace{-2.25pt}r_{(u,v)}(t)\hspace{-2.25pt}+\hspace{-2.25pt}r_{(v,u)}(t)\right)^2\hspace{-3pt}\bigg] \nonumber \\
    &\overset{*}{\leq} \mathbb{E}_{\xs}\bigg[ |E|\bigg(\sum_{i \neq j \in V} a_{ij}(t)+1\bigg)^2 \bigg] \nonumber \\
    & \overset{**}{=} \mathbb{E}\bigg[ |E|\bigg(\sum_{i \neq j \in V} a_{ij}(t)+1\bigg)^2 \bigg] \nonumber \\
    &=|E|\bigg(\sum_{i \neq j \in V}\sigma_{ij}^2+\bigg(\sum_{i \neq j\in V} \lambda_{ij}\bigg)^2+1+2\hspace{-4pt}\sum_{i \neq j \in V} \lambda_{ij}\bigg), \label{eq: proofoftheorem_upperbound}
\end{align}
where $(*)$ follows as the demand routed through any edge at time $t$ is less than or equal to the total incoming demand at time $t$. Also, by the Algorithm 1, $r_{(u,v)} \leq 1$. Next, $(**)$ follows as the incoming demand is independent of the queue lengths. Thus, the first term in \eqref{vsim1} is bounded. Let us denote \eqref{eq: proofoftheorem_upperbound} by a constant $B_1$.  Now we have,
\begin{align*}
     \lefteqn{\mathbb{E}_{\xs}[\Delta V_1]-T_1(\xs)} \\
     \leq{}&  2\mathbb{E}_{\xs}\bigg[\sum_{(u,v) \in E} z_{(u,v)}(t)\left(y_{(u,v)}(t)-y_{(v,u)}(t)-r_{(u,v)}(t) \right. \nonumber \\
      &\left. +r_{(v,u)}(t)\right)\bigg] \nonumber \\
        \overset{*}{\leq}{}&4\mathbb{E}_{\xs}\bigg[\sum_{(u,v) \in E} z_{(u,v)}(t)\bigg(\sum_{i \neq j \in V}\sum_{p \in P_{ij}}x_p(t)\mathbbm{1}_{(u,v) \in p}  \nonumber \\
         & -r_{(u,v)}(t)\bigg)\bigg] \\
            ={}&4\mathbb{E}_{\xs}\bigg[\sum_{i \neq j \in V}\sum_{p \in P_{ij}}x_p(t)\sum_{(u,v)\in p}z_{(u,v)}(t)\bigg] \nonumber\\
            &-4\sum_{(u,v) \in E}z_{(u,v)}(t)\mathbb{E}_{\xs}[r_{(u,v)}(t)] \numberthis \label{eq: V1_drift} 
    \end{align*}
 where $(*)$ follows from the fact that $z_{(u,v)}(t)=-z_{(v,u)}(t)$ and substituting $y_{(u,v)}$ using \eqref{y}. Now we will calculate the drift of $V_2$ and then add it with \eqref{eq: V1_drift} to get the drift of $V$. 
 
We will now calculate the drift of $V_2(\xs)$. We have:

\begin{align*}
    \lefteqn{\mathbb{E}_{\xs}[\Delta V_2]} \\
    = {}& \mathbb{E}_{\xs}\bigg[\sum_{(u,v) \in E}\coeff{2}\bigg(\left(q_{(u,v)}(t+1)+q_{(v,u)}(t+1)\right)^2\nonumber \\
    &-\left(q_{(u,v)}(t)+q_{(v,u)}(t)\right)^2\bigg)\bigg]\\     
    = {}& \sum_{(u,v) \in E}\coeff{2}\bigg(\mathbb{E}_{\xs}\big[\big(y_{(u,v)}(t)
    +y_{(v,u)}(t)-2s_{(u,v)}(t) \nonumber \\
    &-r_{(u,v)}(t)-r_{(v,u)}(t)\big)^2\big]+2\mathbb{E}_{\mathbf{q}}\big[(q_{(u,v)}(t)+q_{(v,u)}) \times \nonumber \\
    &\left(y_{(u,v)}(t)+\hspace{-2pt}y_{(v,u)}(t)\hspace{-2pt}-\hspace{-2pt}2s_{(u,v)}(t)-\hspace{-2pt}r_{(u,v)}(t)-\hspace{-2pt}r_{(v,u)}(t)\right)\big]\bigg). \numberthis \label{vsim}
\end{align*}
We will now show that the first term in \eqref{vsim} is bounded. We have:
\begin{align}
    T_2(\xs)={}&\sum_{(u,v) \in E}\coeff{2}\mathbb{E}_{\xs}\bigg[\big(y_{(u,v)}(t)
    +y_{(v,u)}(t) \nonumber \\&-2s_{(u,v)}(t)-r_{(u,v)}(t)-r_{(v,u)}(t)\big)^2\bigg] \nonumber \\
     \overset{*}{\leq}{}& \frac{|E|\delta}{2\alpha c_{max}} \mathbb{E}_{\xs}\bigg[\bigg(\sum_{i\neq j \in V} a_{ij}(t)\bigg)^2\bigg] \nonumber \\
    ={}&\frac{|E|\delta}{2\alpha c_{max}}\bigg(\sum_{i\neq j \in V}\sigma_{ij}^2+\bigg(\sum_{i\neq j \in V} \lambda_{ij}\bigg)^2\bigg), \label{eq: proofoftheorem_upperboundq}
\end{align}
    where $(*)$ follows as $s_{(u,v)}(t)$ and $\mathbf{r}(t)$ are non negative and the demand routed through any edge at time $t$ is less than or equal to the total incoming demand at time $t$. Thus, the first term in \eqref{vsim} is bounded. Let us denote \eqref{eq: proofoftheorem_upperboundq} by a constant $B_2$.  Now we have,
    \begin{align*}
     \lefteqn{\mathbb{E}_{\xs}[\Delta V_2]-T_2(\xs)} \\ 
     \leq {}& \mathbb{E}_{\xs}\bigg[\sum_{(u,v) \in E}\coeff{}\bigg( (q_{(u,v)}(t)+q_{(v,u)}(t))\times \nonumber \\
     &\left(y_{(u,v)}(t)+y_{(v,u)}(t)-2s_{(u,v)}(t)-\hspace{-2pt}r_{(u,v)}(t)-\hspace{-2pt}r_{(v,u)}(t)\right)\hspace{-3pt}\bigg)\hspace{-2pt}\bigg]  \nonumber \\
         \overset{*}{\leq} {}& 2\mathbb{E}_{\xs}\bigg[\sum_{(u,v) \in E}\coeff{}\bigg( \big(q_{(u,v)}(t)+q_{(v,u)}(t)\big) \times \nonumber \\
         &\bigg(\sum_{i \neq j \in V}\sum_{p \in P_{ij}}x_p(t) \mathbbm{1}_{(u,v) \in p}-s_{(u,v)}(t)-r_{(u,v)}(t)\bigg)\bigg)\bigg]   \\
            ={}&2\mathbb{E}_{\xs}\bigg[\sum_{i \neq j \in V}\sum_{p \in P_{ij}}x_p(t)\sum_{(u,v)\in p}\coeff{}(q_{(u,v)}(t) \nonumber \\
            &+q_{(v,u)}(t))\bigg]-2\sum_{(u,v) \in E}\coeff{}\bigg(2\mathbb{E}_{\xs}[s_{(u,v)}(t)]q_{(u,v)}(t) \nonumber \\
          &-\big(q_{(u,v)}(t)+q_{(v,u)}(t)\big)\mathbb{E}_{\xs}[r_{(u,v)}(t)]\bigg) \numberthis \label{eq: V2_drift}   
\end{align*}
 where $(*)$ follows by substituting $y_{(u,v)}$ using \eqref{y}. Now, we will add \eqref{eq: V1_drift} with \eqref{eq: V2_drift} and use Algorithm \ref{alg2} to upper bound the drift of $V$. By defining $B\overset{\Delta}{=}B_1+B_2$ and $T(\xs)\overset{\Delta}{=}T_1(\xs)+T_2(\xs)$ we have,
          \begin{align*}
             \lefteqn{\mathbb{E}_{\xs}[\Delta V]-T(\xs)} \\
             \leq  {}&  4\sum_{i\neq j \in V}\lambda_{ij}\min_{p \in P_{ij}}\bigg\{\sum_{(u,v) \in p}\big(z_{(u,v)}(t)+\coeff{2}(q_{(u,v)}(t)\nonumber  \\
            & \hspace{-15pt}+q_{(v,u)}(t))\big)\bigg\}-4\sum_{(u,v) \in E}\bigg(\coeff{}\mathbb{E}_{\xs}[s_{(u,v)}(t)]q_{(u,v)}(t) \nonumber \\
            &\hspace{-15pt}-\delta\big(z_{(u,v)}(t)\hspace{-2pt}+\hspace{-2pt}\coeff{2}(q_{(u,v)}(t)+q_{(v,u)}(t))\big)\mathbbm{1}_{q_{(u,v)}(t)>M}\hspace{-2pt}\bigg). \numberthis \label{dri}
    \end{align*}
    The last term in \eqref{dri} appears due to the $\delta$-on chain rebalancing for $q_{(u,v)}$ greater than $M$. 
    
    Now we will use the definition of the capacity region $\mathbb{C}$ and bound the first term in \eqref{dri}. Using Lemma 1 we have,
    \begin{align}
        \lefteqn{\mathbb{E}_{\xs}[\Delta V]}\nonumber \\
        \leq {}& T(\xs)\hspace{-2pt}+\hspace{-2pt}4\delta\sum_{(u,v) \in E} q_{(u,v)}(t)\bigg(\frac{1}{\alpha}-\frac{\mathbb{E}_{\xs}[s_{(u,v)}(t)]}{\alpha c_{(u,v)}}-\mathbbm{1}_{q_{(u,v)}>M} \nonumber \\
        &+\mathbbm{1}_{q_{(v,u)}>M}-\frac{\delta}{2\alpha c_{(u,v)}}(\mathbbm{1}_{q_{(u,v)}>M}+\mathbbm{1}_{q_{(v,u)}>M})\bigg). \label{eq: proof-drift}
    \end{align}
    In order to use the Foster-Lyapunov Theorem, we want to show that this drift is negative outside a finite set $\mathcal{B}$.
    It is not difficult to see that when $q_{(u,v)}>M>c_{\max}$, then the coefficient of $q_{(u,v)}$ in \eqref{eq: proof-drift} is less than or equal to $-\coeff{2}$, by considering the two cases when $q_{(v,u)}>M$ and $q_{(v,u)}\leq M$.

    Consider the following two cases: \newline
    \textbf{Case 1} [$q_{(u,v)}>M$ and $q_{(v,u)}>M$]: The coefficient of $q_{(u,v)}$ in \eqref{eq: proof-drift} becomes $-\coeff{}$ as $s_{(u,v)}=c_{(u,v)}$ by the definition of $s_{(u,v)}$. \newline
    \textbf{Case 2} [$q_{(u,v)}>M$ and $q_{(v,u)}<M$]: The coefficient of $q_{(u,v)}$ in \eqref{eq: proof-drift} becomes $\frac{1}{\alpha}-1-\frac{s_{(u,v)}(t)}{\alpha c_{(u,v)}}-\coeff{2}$ which is less than or equal to $-\coeff{2} \leq 0$ as $\alpha>1$.
    
    The maximum value of the coefficient of $q_{(u,v)}$ for any $(u,v) \in E$ is $4\delta(\frac{1}{\alpha}+1)<8\delta$, thus the maximum value of the RHS of \eqref{eq: proof-drift} will be $B+8\delta M |E|$ as the coefficient of $q_{(u,v)}$ for any $(u,v) \in E$ is positive only when $q_{(u,v)} \leq M$ and $T(\xs) \leq B$ for any $\xs$. Thus, for at least one edge say $(\widetilde{u},\widetilde{v})$ if we have $q_{(\widetilde{u},\widetilde{v})} \geq (8\delta M|E|+B)\frac{\alpha c_{(\widetilde{u},\widetilde{v})}}{2\delta^2}+M$, the RHS in \eqref{eq: proof-drift} will become less than or equal to $-2M\frac{\delta^2}{\alpha c_{(\widetilde{u},\widetilde{v})}}$.

    \begin{align}
        \lefteqn{\mathbb{E}_{\xs}[\Delta V]} \nonumber \\  \leq{}&B-(8\delta M|E|+B)-2M\frac{\delta^2}{\alpha c_{(u,v)}}+4\delta \hspace{-12pt}\sum_{(u,v) \in E/(\widetilde{u},\widetilde{v})}\hspace{-10pt}q_{(u,v)}(t)\times \nonumber \\
        &\bigg(\frac{1}{\alpha}-\frac{s_{(u,v)}(t)}{\alpha c_{(u,v)}}-\mathbbm{1}_{q_{(u,v)}>M}+\mathbbm{1}_{q_{(v,u)}>M}-\frac{\delta}{2\alpha c_{(u,v)}}\times \nonumber \\
        &(\mathbbm{1}_{q_{(u,v)}>M}+\mathbbm{1}_{q_{(v,u)}>M})\bigg), \nonumber \\
        \leq{}& -2M\frac{\delta^2}{\alpha c_{(u,v)}} < 0.
    \end{align}
    
    So, we can conclude that the drift of $V(q)$ is negative outside the finite set $\mathcal{B}$ which is defined as:
    
    \begin{align}
        \mathcal{B}\overset{\Delta}{=}{}&\bigg\{\mathbf{q} \in \mathbb{Z}_{+}^{|E|}: \mathbf{q} \leq \frac{4\alpha M|E|\mathbf{c}}{\delta}+\frac{\alpha B \mathbf{c}}{2\delta^2}+M\mathbf{1}\bigg\}. \nonumber
    \end{align}
   Note that the operator ``$\leq$'' is applied component wise and $\mathbf{1}$ is a vector of 1's. As the drift of $V(\mathbf{q})$ is negative outside the finite set $\mathcal{B}$, the DTMC $\{\mathbf{q}(t): t \geq 0\}$ is positive recurrent. 

The rate of on chain rebalancing under the algorithm by $\delta$ and $M$ is
$\sum_{(u,v)\in E}\mathbb{E}[r_{(u,v)}]P[q_{(u,v)}>M] \leq \delta |E|$, which is $O(\delta)$.

 Thus, given any $\epsilon>0$, we can set $\delta=\frac{\epsilon}{|E|}$ such that the rate of on chain rebalancing is less than or equal to $\epsilon$. Thus, the family of algorithms in Algorithm \ref{alg2} is strongly stable, and so a part of the theorem follows.
 
 We will now show that the expected value of the queue length is bounded. Upper bounding the Right hand side of \eqref{eq: proof-drift} considering the two cases when $q>M$ and $q\leq M$ gives:
 \begin{align}
     \lefteqn{\mathbb{E}_{\xs}[\Delta V]-T(\xs)} \nonumber \\ 
     \leq{}& 4\delta \sum_{(u,v) \in E} q_{(u,v)}(t)\bigg(2\mathbbm{1}_{q_{(u,v)}\leq M}-\coeff{2}\mathbbm{1}_{q_{(u,v)}>M}\bigg) \nonumber \\
     ={}& 4\delta \sum_{(u,v) \in E} \hspace{-3pt} q_{(u,v)}(t)\bigg((2+\coeff{2})\mathbbm{1}_{q_{(u,v)}\leq M}-\coeff{2}\bigg). \nonumber
 \end{align}
 By moment bound theorem \cite{HajekComm}, we have
 \begin{align}
     \mathbb{E}[T(\xs)]+4\delta \sum_{(u,v) \in E}&(2+\coeff{2})\mathbb{E}\left[q_{(u,v)}(t)\mathbbm{1}_{q_{(u,v)(t)\leq M}}\right] \nonumber \\
     \geq{}& \frac{2\delta^2}{\alpha}\sum_{(u,v) \in E} \mathbb{E}\left[\frac{q_{(u,v)}(t)}{c_{(u,v)}}\right], \label{eq: momentbound}
 \end{align}
where the expectation is with respect to the stationary distribution of the DTMC operating under Algorithm \ref{alg2} parametrized by $\delta, M$ and $\alpha$. Simplifying \eqref{eq: momentbound} by using the inequalities $\mathbb{E}\left[q_{(u,v)}(t)\mathbbm{1}_{q_{(u,v)(t)\leq M}}\right] \leq M$ and $\sum_{(u,v) \in E}\mathbb{E}[\frac{q_{(u,v)}(t)}{c_{(u,v)}}] \geq \frac{1}{c_{max}}\sum_{(u,v) \in E} \mathbb{E}[q_{(u,v)}(t)]$ we get,
 \begin{align}
     \sum_{(u,v) \in E} \mathbb{E}[q_{(u,v)}(t)] \leq{}& \frac{\mathbb{E}[T(\xs)] \alpha c_{max}}{2\delta^2}+\frac{4M\alpha c_{max}|E|}{\delta}\nonumber \\
     &+\frac{M c_{max} |E|}{c_{min}} \label{eq: proof1boundonq}
 \end{align}
Now, we will find the expression of $\mathbb{E}[T(\xs)]$ by bounding the quadratic terms. We know that $\{\mathbf{q}(t):t \in \mathbb{Z}_{+}\}$ is positive recurrent and  $\mathbb{E}[\mathbf{q}]$ is finite, thus under the stationary distribution of the DTMC, we have
\begin{align}
    &\mathbb{E}[z_{(u,v)}(t+1)-z_{(u,v)}(t)] \nonumber \\
    ={}&\mathbb{E}[y_{(u,v)}(t)-y_{(v,u)}(t)-r_{(u,v)}(t)+r_{(v,u)}(t)]=0 \label{eq: proof1stationarityz}
\end{align}
We will first calculate the expression of $\mathbb{E}[T_1(\xs)]$ as follows:
\begin{align}
    \lefteqn{\mathbb{E}[T_1(\xs)]} \nonumber \\
    ={}& \hspace{-2pt} \sum_{(u,v) \in E}\hspace{-2pt} \mathbb{E}[(y_{(u,v)}(t)-y_{(v,u)}(t)-r_{(u,v)}(t)+r_{(v,u)}(t))^2] \nonumber \\
    \overset{\alpha}{=}{}&\hspace{-2pt}\sum_{(u,v) \in E} \mathrm{Var}[y_{(u,v)}(t)-y_{(v,u)}(t)-r_{(u,v)}(t)+r_{(v,u)}(t)] \nonumber \\
    \overset{\beta}{\leq}{}& \sum_{(u,v) \in E}\bigg( 4\mathrm{Var}[y_{(u,v)}(t)]+4\mathrm{Var}[r_{(u,v)}(t)]+ \nonumber \\
    &\mathrm{Cov}[y_{(u,v)}(t)-y_{(v,u)}(t),r_{(u,v)}(t)-r_{(v,u)}(t)]\bigg) \nonumber \\
    \overset{\gamma}{\leq}{}& \sum_{(u,v) \in E}\bigg(4\mathrm{Var}[y_{(u,v)}(t)]+4\mathrm{Var}[r_{(u,v)}(t)]+ \nonumber \\
    & \mathrm{Var}[y_{(u,v)}(t)\hspace{-1.4pt}-\hspace{-1.4pt}y_{(v,u)}(t)]\hspace{-2pt}+\hspace{-2pt}\mathrm{Var}[r_{(u,v)}(t)\hspace{-1.4pt}-\hspace{-1.4pt}r_{(v,u)}(t)]\hspace{-2pt}\bigg) \nonumber \\
    \leq{}& 8 \sum_{(u,v) \in E} \bigg(\mathrm{Var}[y_{(u,v)}(t)]+\mathrm{Var}[r_{(u,v)}(t)]\bigg) \nonumber \\
    \overset{\xi}{\leq}{}& 8\sum_{(u,v) \in E} \bigg( \sum_{i \neq j \in V} \sigma^2_{ij}+\delta\bigg) \nonumber \\
    ={}& 8|E|\bigg(\sum_{i \neq j \in V} \sigma^2_{ij}+\delta\bigg) \label{eq: proof1boundB1},
\end{align}
where $\alpha$ follows from \eqref{eq: proof1stationarityz}, $\beta$ and $\gamma$ follows from the equations $\mathrm{Var}[A+B]=\mathrm{Var}[A]+\mathrm{Var}[B]+2\mathrm{Cov}[A,B]$ and $\mathrm{Cov}[A,B] \leq \sqrt{\mathrm{Var}[A]\mathrm{Var}[B]} \leq (\mathrm{Var}[A]+\mathrm{Var}[B])/2$. Now, $\xi$ follows from the fact that the amount of demand routed through each edge at any time $t$ is less than or equal to the total incoming demand at time $t$ and the i.id assumption of $\mathbf{a}$ across vertices and time where $\sigma_{ij}^2$ is the variance of the incoming demand.

Similarly, we will now calculate the expression of $\mathbb{E}[T_2(\xs)]$. We will omit some steps as they are repetitive and directly write:
\begin{align}
    \mathbb{E}[T_2(\xs)]\leq{}& 2\sum_{(u,v)\in E} \bigg( \mathrm{Var}[y_{(u,v)}(t)+y_{(v,u)}(t) \nonumber \\
    &-r_{(u,v)}(t)-r_{(v,u)}(t)]+2\mathrm{Var}[s_{(u,v)}]\bigg) \nonumber \\
    \overset{*}{\leq}& 16|E|\bigg(\sum_{i \neq j \in V} \sigma^2_{ij}+\delta\bigg)+4\sum_{(u,v) \in E} c_{(u,v)}^2, \label{eq: proof1boundB2}
\end{align}
where $(*)$ follows from \eqref{eq: proof1boundB1} and $s_{(u,v)} \leq c_{(u,v)}$. Now, we can use \eqref{eq: proof1boundB1} and \eqref{eq: proof1boundB2} to write the expression of $\mathbb{E}[T(\xs)]$ as follows:
\begin{align}
    \mathbb{E}[T(\xs)]\leq24|E|\sum_{i \neq j \in V} \sigma_{ij}^2+4\hspace{-2pt}\sum_{(u,v) \in E} c_{(u,v)}^2+24|E|\delta.
\end{align}
Note that the bound on $\mathbb{E}[T(\xs)]$ is loose and can be tighten considerably. Especially, if the network topology is known, we can have a tighter bound on $\mathrm{Var}[\mathbf{y}]$.

By substituting the expression of $\mathbb{E}[T(\xs)]$ in \eqref{eq: proof1boundonq}, the proof follows.
\end{proof}
\subsection{Proof of Lemma \ref{lemma}}
\begin{proof}
We start with the feasible Linear program (LP) for $\vlam \in C$ as,
\begin{align}
     \max_{x_p \in \mathbb{R}^{|P|}_{+}} & \ 0  \nonumber 
\end{align}
subject to \eqref{prop1}, \eqref{prop2} and \eqref{prop3}. Taking $\zeta_{ij}, \beta_{(u,v)}$ and $\gamma_{(u,v)}$ as the dual variables of \eqref{prop1}, \eqref{prop2} and \eqref{prop3} respectively, the dual of the above LP is,
\begin{align}
    \min \sum_{i \neq j \in V} \lambda_{ij}\zeta_{ij}+2\sum_{(u,v) \in E} c_{(u,v)}&\beta_{(u,v)} \label{eq: obj-dual}
\end{align}
subject to,
\begin{align}
    \sum_{(u,v) \in p}(\beta_{(u,v)}&+\beta_{(v,u)}+\gamma_{(u,v)}-\gamma_{(v,u)})  \nonumber \\
    &\geq  -\zeta_{ij} \  \forall p \in P_{ij} \ \forall i \neq j \in V, \label{constrant1} \\
    \beta_{(u,v)} &\geq 0 \ \forall (u,v) \in E. \label{constraint2}
\end{align}
Note that the constraints in the primal LP defines the capacity region $\mathcal{C}$. Therefore, since  $\vlam \in C$, we have that the primal is feasible. The dual is feasible as one trivial feasible solution is all the variables are zero. By strong duality, the dual should have an optimal value equal to 0. Thus we have,
\begin{align}
    \sum_{i \neq j \in V} \lambda_{ij}\zeta_{ij}+2\sum_{(u,v) \in E} c_{(u,v)}\beta_{(u,v)} &\geq 0 \label{dualmagic}
\end{align}
for any $\zeta_{ij}, \beta_{(u,v)}$ and $\gamma_{(u,v)}$ subject to \eqref{constrant1} and \eqref{constraint2}. We pick  $\zeta_{ij}= -\min_{p \in P_{ij}}\sum_{(u,v) \in p}(\beta_{(u,v)} +\beta_{(v,u)}+\gamma_{(u,v)}-\gamma_{(v,u)})$ for all $i \neq j \in V$, to get
\begin{align}
      2\sum_{(u,v) \in E} c_{(u,v)}\beta_{(u,v)} \geq &\sum_{i\neq j \in V}\lambda_{ij}\min_{p \in P_{ij}}\sum_{(u,v) \in p}(\beta_{(u,v)} \nonumber \\
      &+\beta_{(v,u)}+\gamma_{(u,v)}-\gamma_{(v,u)}) \nonumber,
\end{align}
for all $\beta_{(u,v)} \geq 0$ and $\gamma_{(u,v)} \in \mathbb{R}$. By taking $\beta_{(u,v)}=\coeff{}q_{(u,v)}$ and $\gamma_{(u,v)}=z_{(u,v)} $, and noting that $z_{(u,v)}=-z_{(v,u)}$ we have the lemma.
\end{proof}
\end{document}